\documentclass[12pt,letterpaper]{article}
\usepackage{threeparttable}
\usepackage[a4paper, total={7in, 10in}]{geometry}

\usepackage{helvet}
\usepackage{authblk}
\usepackage{graphicx}
\usepackage{hyperref}
\usepackage{amsmath,amssymb,amsfonts,amsthm,mathrsfs}
\usepackage[super,comma,sort&compress]{natbib} 
\usepackage{rotating}
\usepackage{multirow}
\usepackage{booktabs}
\usepackage{algorithm}
\usepackage{algorithmicx}
\usepackage{algpseudocode}
\usepackage{listings}
\usepackage{subcaption}
\usepackage{tabularx}
\usepackage{placeins} 
\usepackage{xcolor}

\graphicspath{{images/}} 

\makeatletter
\renewcommand{\maketitle}{\bgroup\setlength{\parindent}{0pt}
\begin{flushleft}
  \textbf{\large \@title} \par \vspace{1em}
  \@author
\end{flushleft}\egroup}
\makeatother

\title{Aortic Valve Disease Screening from PPG via Physiology-Guided Self-Supervised Learning}

\author[1,2,$\dagger$]{Jiaze Wang}
\author[3,$\dagger$]{Qinghao Zhao}
\author[1,$\dagger$]{Zizheng Chen}
\author[1]{Zhejun Sun}
\author[4]{Deyun Zhang}
\author[1,5,*]{Yuxi Zhou}
\author[2,6,7,8,*]{Shenda Hong}

\affil[1]{Department of Computer Science, Tianjin University of Technology, Tianjin, China}
\affil[2]{National Institute of Health Data Science, Peking University, Beijing, China}
\affil[3]{Department of Cardiology, Peking University People’s Hospital, Beijing, China}
\affil[4]{HeartVoice Medical Technology, Hefei, China}
\affil[5]{DCST, BNRist, RIIT, Institute of Internet Industry, Tsinghua University, Beijing, China}
\affil[6]{Institute of Medical Technology, Peking University Health Science Center, Beijing, China}
\affil[7]{Institute for Artificial Intelligence, Peking University, Beijing, China}
\affil[8]{State Key Laboratory of Vascular Homeostasis and Remodeling, NHC Key Laboratory of Cardiovascular Molecular Biology and Regulatory Peptides, Peking University, Beijing, China}

\affil[$\dagger$]{These authors contributed equally}
\affil[*]{Correspondence: joy\_yuxi@pku.edu.cn, hongshenda@pku.edu.cn}
\begin{document}

\maketitle
\section*{ABSTRACT}
Aortic valve disease (AVD) represents a major public health burden, while its diagnosis relies on echocardiography, which is limited by cost and specialist expertise, restricting scalable screening and risk stratification. Existing portable sensing modalities are constrained by indirect representations or acquisition dependencies. In this context, photoplethysmography (PPG), a widely available optical signal capturing peripheral hemodynamic dynamics, provides a scalable physiological measurement. However, the scarcity of clinically labeled PPG data severely constrains the development of effective data-driven models. To address this limitation, we propose Physiology-Guided Self-Supervised Learning (PG-SSL), leveraging approximately 170,000 unlabeled UK Biobank PPG recordings. PG-SSL constructs physiologically derived pseudo-labels based on clinically motivated waveform phenotypes associated with aortic stenosis (AS) and aortic regurgitation (AR), enabling large-scale pretraining without AVD-specific labels. Following fine-tuning on a small labeled cohort, the model achieved AUROCs of 0.8025 for AS and 0.7669 for AR. Further analyses demonstrated robustness under clinical confounding and covariate-balanced evaluation, as well as significant longitudinal associations with incident AVD events. This study demonstrates the feasibility of PG-SSL for leveraging large-scale unlabeled physiological signals under clinically labeled data-scarce conditions. The proposed approach provides a useful strategy for improving low-cost PPG-based screening and risk enrichment for clinically recognized AVD.

\section*{KEYWORDS}
Aortic Valve Disease, Photoplethysmography, Physiology-Guided Self-Supervised Learning, Representation Learning, Disease Screening, UK Biobank

\section*{INTRODUCTION}
Valvular heart disease (VHD) stands as a primary cause of global cardiovascular morbidity and mortality, with its burden escalating dramatically alongside the aging of the world's population~\cite{coffey2021global}. Among all forms of VHD, aortic valve disease (AVD), which predominantly includes aortic stenosis (AS) and aortic regurgitation (AR), represents the most life-threatening category, accounting for a staggering 61\% of all VHD-related deaths~\cite{aluru2022valvular}. However, a stark clinical reality persists: despite the severe consequences of AVD, its early recognition remains challenging. The disease progression is typically insidious, with patients remaining largely asymptomatic until the optimal window for intervention has passed. Once symptoms emerge, the average life expectancy can plummet to as little as two to three years~\cite{thoenes2018patient}. Therefore, there is a growing need for scalable and low-cost screening strategies capable of enriching high-risk individuals who may benefit from further cardiovascular evaluation during the asymptomatic phase of AVD.

From the clinical gold standard to emerging portable sensors, existing solutions face fundamental bottlenecks in meeting the requirements of convenience, reliability, and physiological relevance necessary for large-scale screening. While imaging-based gold standards like transthoracic echocardiography (TTE) offer diagnostic precision~\cite{praz2025esc}, their heavy reliance on specialized personnel and dedicated facilities makes them impractical for widespread deployment. Emerging portable technologies based on chest wall vibrations, such as digital phonocardiography (PCG), seismocardiography (SCG), and gyrocardiography (GCG)~\cite{zhao2024deep, khade2021machine, erin2024performance, saraf2019fully}, rely on stable and precise physical contact with the skin, making it difficult for non-expert users to ensure data quality and reproducibility. The more widely used portable electrocardiogram (ECG)~\cite{cohen2021electrocardiogram, sawano2022deep}, in turn, faces a fundamental limitation of informational indirectness. AVD is inherently a mechanical disorder, whereas the ECG primarily captures electrical activity that may reflect downstream and non-specific consequences of long-term hemodynamic stress, potentially limiting its sensitivity to subtle valvular hemodynamic abnormalities. In summary, existing portable solutions are either constrained by physical acquisition dependencies or limited in their ability to directly capture peripheral hemodynamic dynamics, leaving a clear need for a sensing modality that is both easily deployable and physiologically informative for large-scale screening.

To address these challenges, there is an urgent need for a sensing technology that can capture hemodynamic dynamics while being amenable to large-scale deployment. Photoplethysmography (PPG) presents a highly compelling solution. As an optical sensing technology already integrated into hundreds of millions of consumer wearable devices worldwide, PPG possesses a strong hardware foundation for scalable cardiovascular screening~\cite{pereira2020photoplethysmography, nie2024ppg}. More critically, the core advantage of PPG lies in its close association with peripheral hemodynamic dynamics. By measuring pulsatile blood volume changes in peripheral tissues, it captures morphological characteristics related to the central arterial pressure wave~\cite{allen2007photoplethysmography}, offering a non-invasive window into cardiovascular hemodynamic alterations associated with valvular dysfunction. Specifically, obstructed ejection in AS and abnormal diastolic pressure dynamics in AR may influence the contour, amplitude, and timing characteristics of the PPG waveform. Indeed, the feasibility of using PPG morphological analysis to assess complex cardiovascular states, such as blood pressure, arterial stiffness, and atrial fibrillation, has been extensively validated~\cite{elgendi2019use}. While nascent studies have explored the use of PPG for detecting isolated AS~\cite{yang2024aortic, kang2025pass}, its application to AR screening remains relatively underexplored. Furthermore, unified AI frameworks capable of simultaneously screening for both AS and AR from PPG signals remain limited. Therefore, given its scalability and physiological relevance, PPG provides a promising foundation for low-cost AVD screening and risk-enrichment strategies.

Despite the unique physiological perspective PPG offers for AVD screening, translating this potential into accurate and robust algorithms faces a dual challenge spanning feature representation and data availability. First, waveform alterations potentially associated with valvular dysfunction are often relatively subtle and are frequently confounded by morphological variations arising from individual physiological differences such as age and arterial stiffness~\cite{millasseau2006contour}. This inherent signal complexity demands models capable of capturing subtle disease-associated waveform patterns despite substantial physiological variation across individuals~\cite{karimpour2023photoplethysmography}. This feature extraction challenge is further amplified by the scarcity of high-quality supervised data. In existing public datasets and large cohort studies, the number of PPG samples aligned with clinically recognized AS/AR labels or echocardiographic assessments remains limited. This scarcity of labeled data, a widely recognized bottleneck in medical artificial intelligence~\cite{krishnan2022self}, substantially constrains the effective training of complex deep learning models. Consequently, the field confronts a central paradox: the need for powerful models capable of learning subtle physiological representations starkly contrasts with the lack of large-scale, high-quality labeled data required to train them. This paradox motivates the exploration of representation learning paradigms beyond conventional supervised learning frameworks.

To address the dual challenges of subtle feature representation and label scarcity, we developed the \textbf{P}hysiology-\textbf{i}nformed \textbf{L}earning \textbf{A}rchitecture (PiLA), a physiology-informed representation learning framework built upon Physiology-Guided Self-Supervised Learning (PG-SSL). PG-SSL is designed as a physiology-guided extension of the conventional pretrain--fine-tune self-supervised learning paradigm for physiological signal analysis. Unlike classical self-supervised learning methods that rely solely on intrinsic data structure, PG-SSL incorporates clinically motivated physiological priors to construct physiology-guided proxy supervisory signals from large-scale unlabeled PPG data, enabling representation learning without manual pathological annotation. Specifically, we formalize established physiological characteristics of AVD---such as the \textit{pulsus tardus} pattern associated with obstructed ejection in AS, and the \textit{water-hammer pulse} (WHP) pattern associated with diastolic regurgitation in AR~\cite{mcgee2021evidence}---into a set of computable PPG morphological phenotypes. These phenotypes are then used to construct a proxy classification task on a large-scale dataset containing nearly 170,000 unlabeled PPG samples. Through this pre-training process, the model is encouraged to learn physiologically relevant waveform representations from large-scale unlabeled data before downstream adaptation using limited labeled AS/AR samples. Building on this foundation, the PiLA framework is subsequently fine-tuned for downstream AVD screening tasks. Using a large-scale cohort dataset, we show that PiLA improves screening performance for AS and AR compared with supervised learning using limited labeled data alone. Beyond screening performance, we further evaluated the longitudinal association, physiological relevance, and robustness of the learned representations across heterogeneous physiological backgrounds. Collectively, these findings suggest that physiology-guided self-supervised representation learning may provide a useful pathway for leveraging unlabeled physiological signals in data-limited cardiovascular screening settings.

\section*{RESULTS}

\subsection{Study Population}

\begin{figure*}[htbp]
  \centering
  \includegraphics[width=0.90\textwidth]{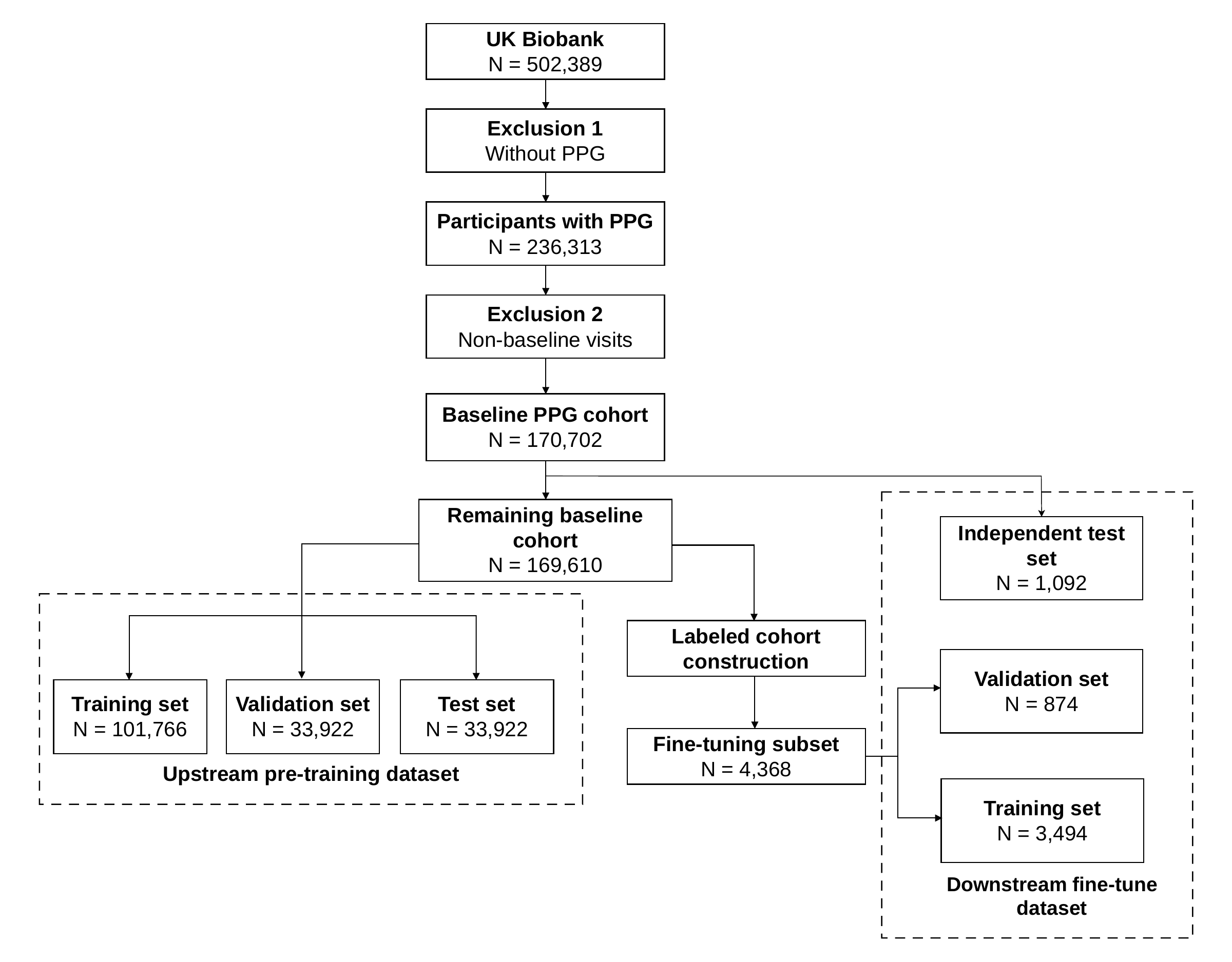}
  \caption{
Data flow chart of cohort construction and data utilization. Starting from 502,389 UK Biobank participants, individuals without available PPG recordings and non-baseline acquisitions were excluded to obtain the baseline PPG cohort. The baseline cohort was subsequently used to construct both the physiology-guided pretraining cohort and the diagnosis-defined downstream cohort for supervised fine-tuning and evaluation.
  }
  \label{fig:dataflow}
\end{figure*}

The study was based on 170,702 baseline photoplethysmography recordings from the UK Biobank. A diagnosis-defined downstream labeled cohort was constructed by combining all identified AS and AR cases with randomly sampled eligible controls, resulting in 5,460 unique participants, including 245 AS-positive participants, 213 AR-positive participants, and 5,083 controls. Among them, 81 participants met criteria for both AS and AR. The downstream cohort was partitioned at the subject level into training, validation, and independent test subsets. After reserving the independent test subset for final evaluation, 169,610 baseline PPG recordings remained available for physiology-guided pretraining. During pretraining, no AS/AR diagnostic labels were used. Participants assigned to the downstream training and validation subsets could contribute to pretraining only through unlabeled PPG waveforms and physiology-derived proxy supervision, whereas AS/AR labels were introduced exclusively during downstream supervised fine-tuning. The overall cohort construction workflow and data utilization strategy are summarized in Fig.~\ref{fig:dataflow}.

Baseline characteristics of the downstream cohort are summarized in Table~\ref{tab:baseline}. Compared with controls, participants with AS or AR exhibited substantial differences across demographic characteristics, hemodynamic measures, metabolic risk factors, and cardiovascular comorbidity profiles. These imbalances motivated subsequent clinical-feature reference analyses and propensity-score-based covariate balancing analyses to evaluate the influence of measurable clinical confounding.

\begin{table}[htbp]
\centering
\caption{\textbf{Baseline characteristics of study cohorts.} Continuous variables are mean $\pm$ SD; categorical are n (\%). P-values compare AS/AR vs Control.}
\label{tab:baseline}

\tiny 
\setlength{\tabcolsep}{1.5pt}

\begin{tabular}{lccccccc}
\toprule
\textbf{Characteristic} & \textbf{Pre-train} & \textbf{Downstream} & \textbf{AS} & \textbf{AR} & \textbf{Control} & \textbf{$P_{AS}$} & \textbf{$P_{AR}$} \\
 & (N=169,610) & (N=5,460) & (N=245) & (N=213) & (N=5,083) & & \\
\midrule

Age (years) & 56.77 $\pm$ 8.16 & 56.87 $\pm$ 8.09 & 62.00 $\pm$ 6.17 & 60.36 $\pm$ 7.09 & 56.55 $\pm$ 8.10 & $<$0.001 & $<$0.001 \\

Sex & & & & & & & \\
\quad Male & 77,759 (45.8\%) & 2,530 (46.3\%) & 151 (61.6\%) & 125 (58.7\%) & 2,308 (45.4\%) & $<$0.001 & $<$0.001 \\

Race & & & & & & & \\
\quad White & 153,666 (90.6\%) & 4,955 (90.8\%) & 229 (93.5\%) & 194 (91.1\%) & 4,609 (90.7\%) & 0.172 & 0.937 \\
\quad Asian & 6,076 (3.6\%) & 191 (3.5\%) & 7 (2.9\%) & 8 (3.8\%) & 177 (3.5\%) & 0.731 & 0.982 \\
\quad Black & 4,567 (2.7\%) & 132 (2.4\%) & 3 (1.2\%) & 4 (1.9\%) & 126 (2.5\%) & 0.286 & 0.820 \\
\addlinespace

BMI (kg/m$^2$) & 27.46 $\pm$ 4.83 & 27.58 $\pm$ 4.89 & 29.79 $\pm$ 5.61 & 28.31 $\pm$ 4.84 & 27.47 $\pm$ 4.83 & $<$0.001 & 0.015 \\
Systolic BP (mmHg) & 137.82 $\pm$ 18.62 & 138.39 $\pm$ 19.10 & 142.51 $\pm$ 19.08 & 141.83 $\pm$ 21.21 & 138.10 $\pm$ 18.99 & $<$0.001 & 0.012 \\
Diastolic BP (mmHg) & 82.12 $\pm$ 10.06 & 82.12 $\pm$ 10.33 & 78.20 $\pm$ 10.80 & 78.19 $\pm$ 11.36 & 82.39 $\pm$ 10.22 & $<$0.001 & $<$0.001 \\
Heart Rate (bpm) & 68.64 $\pm$ 11.16 & 68.36 $\pm$ 11.32 & 67.28 $\pm$ 12.44 & 67.61 $\pm$ 11.57 & 68.42 $\pm$ 11.24 & 0.164 & 0.317 \\
\addlinespace
Arterial Stiffness (m/s) & 9.34 $\pm$ 4.05 & 9.32 $\pm$ 5.20 & 10.00 $\pm$ 3.10 & 9.53 $\pm$ 3.17 & 9.29 $\pm$ 5.32 & $<$0.001 & 0.294 \\
Reflection Index (\%) & 67.72 $\pm$ 32.21 & 67.96 $\pm$ 30.75 & 66.84 $\pm$ 18.30 & 66.76 $\pm$ 14.50 & 68.07 $\pm$ 31.53 & 0.328 & 0.230 \\
\addlinespace
HbA1c (mmol/mol) & 36.31 $\pm$ 6.89 & 36.35 $\pm$ 6.78 & 39.25 $\pm$ 10.01 & 37.65 $\pm$ 8.84 & 36.19 $\pm$ 6.54 & $<$0.001 & 0.023 \\
Cholesterol (mmol/L) & 5.68 $\pm$ 1.15 & 5.64 $\pm$ 1.15 & 5.19 $\pm$ 1.24 & 5.24 $\pm$ 1.17 & 5.67 $\pm$ 1.14 & $<$0.001 & $<$0.001 \\
\addlinespace
Current Smoker & 17,121 (10.2\%) & 544 (10.0\%) & 21 (8.6\%) & 21 (10.0\%) & 510 (10.1\%) & 0.514 & 1.000 \\
Daily Alcohol Intake & 34,152 (20.2\%) & 1,130 (20.8\%) & 58 (23.8\%) & 39 (18.6\%) & 1,050 (20.7\%) & 0.284 & 0.508 \\
\addlinespace
Hypertension & 64,524 (38.0\%) & 2,182 (40.0\%) & 197 (80.4\%) & 157 (73.7\%) & 1,895 (37.3\%) & $<$0.001 & $<$0.001 \\
Diabetes Mellitus & 16,449 (9.7\%) & 566 (10.4\%) & 63 (25.7\%) & 35 (16.4\%) & 486 (9.6\%) & $<$0.001 & 0.001 \\
Hyperlipidemia & 39,884 (23.5\%) & 1,368 (25.1\%) & 172 (70.2\%) & 116 (54.5\%) & 1,138 (22.4\%) & $<$0.001 & $<$0.001 \\
Chronic Kidney Disease & 6,412 (3.8\%) & 243 (4.5\%) & 49 (20.0\%) & 37 (17.4\%) & 178 (3.5\%) & $<$0.001 & $<$0.001 \\
CAD / MI & 19,593 (11.6\%) & 779 (14.3\%) & 158 (64.5\%) & 110 (51.6\%) & 564 (11.1\%) & $<$0.001 & $<$0.001 \\
Atrial Fibrillation & 12,029 (7.1\%) & 488 (8.9\%) & 106 (43.3\%) & 92 (43.2\%) & 330 (6.5\%) & $<$0.001 & $<$0.001 \\
Stroke & 5,650 (3.3\%) & 207 (3.8\%) & 41 (16.7\%) & 32 (15.0\%) & 151 (3.0\%) & $<$0.001 & $<$0.001 \\
\bottomrule
\end{tabular}
\end{table}

\FloatBarrier

\subsection{Screening Performance}

\begin{figure}[htbp]
\centering

\begin{subfigure}[b]{0.48\textwidth}
    \centering
    \includegraphics[width=\linewidth]{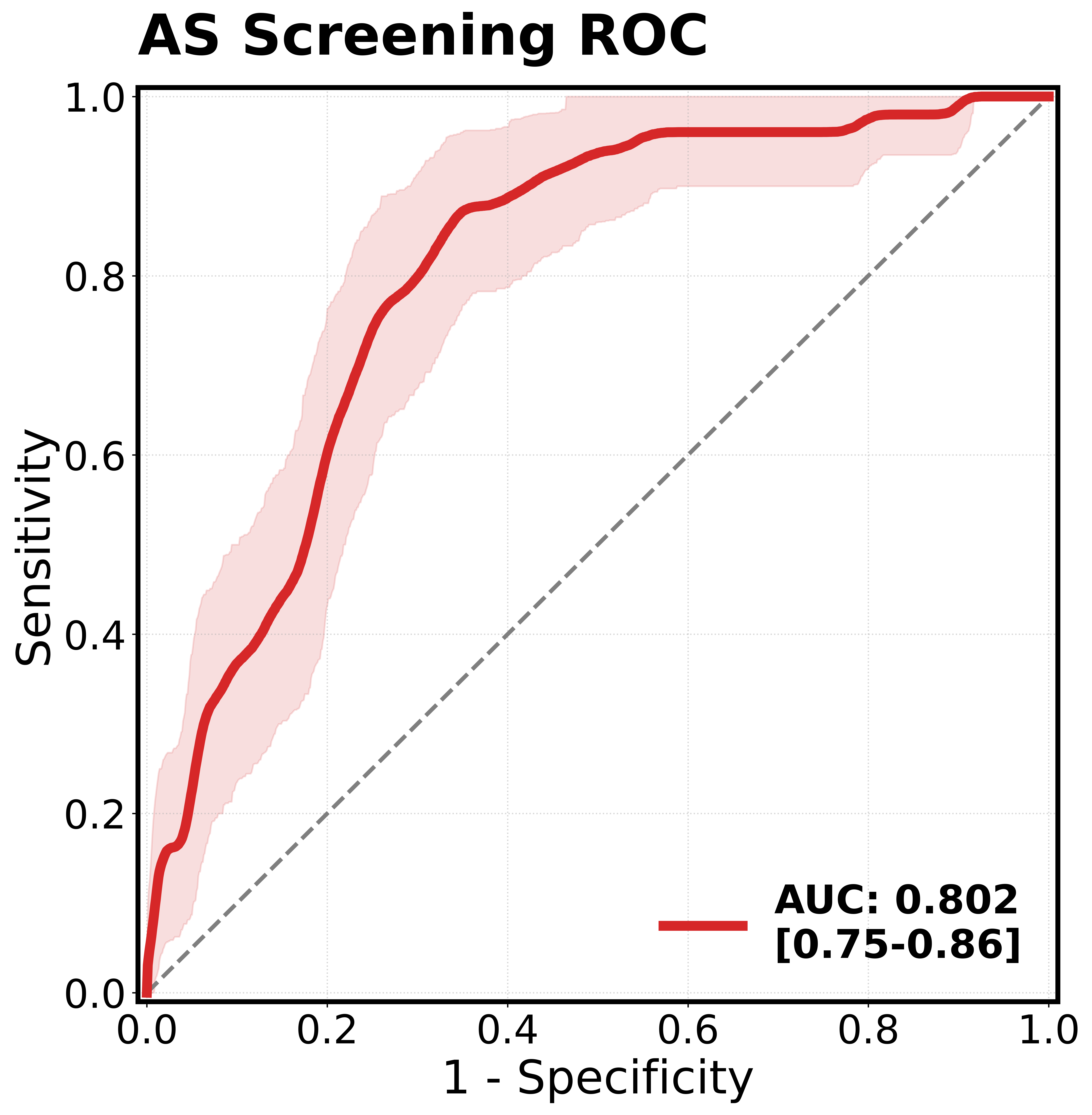}
    \caption{}
    \label{fig:roc_as}
\end{subfigure}
\hfill
\begin{subfigure}[b]{0.48\textwidth}
    \centering
    \includegraphics[width=\linewidth]{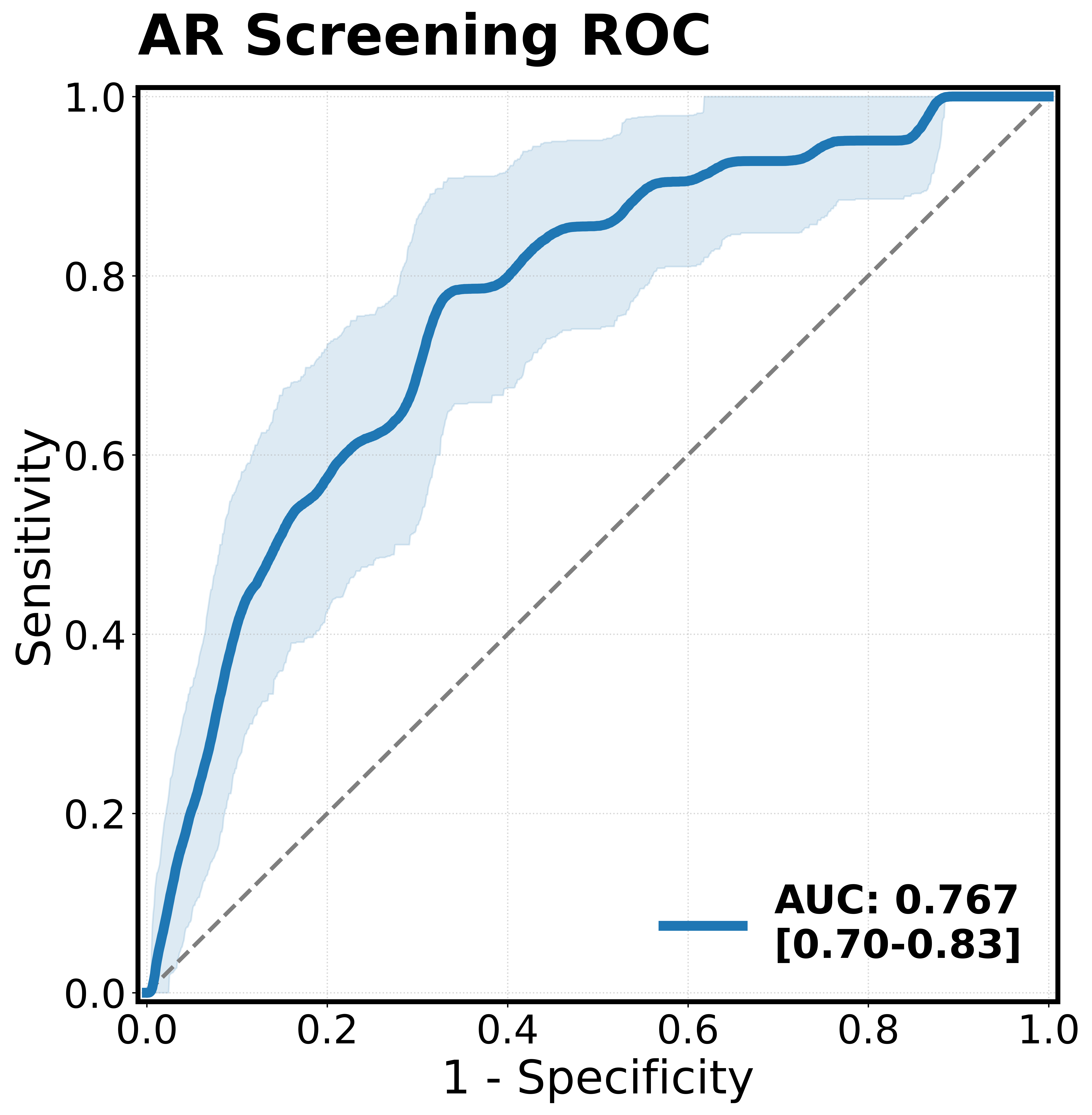}
    \caption{}
    \label{fig:roc_ar}
\end{subfigure}

\caption{
ROC curves of PiLA on the independent test set. a) ROC curve for AS screening. b) ROC curve for AR screening.
}
\label{fig:roc}
\end{figure}

We comprehensively evaluated the screening performance of the PiLA framework on the independent test set. PiLA achieved an Area Under the Receiver Operating Characteristic Curve (AUROC) of 0.8025 for AS and 0.7669 for AR. The Receiver Operating Characteristic (ROC) curves are illustrated in Figure~\ref{fig:roc}. Beyond overall discrimination capability, sensitivity at screening-oriented operating points is also important for risk-enrichment applications. At an operating point corresponding to 70\% specificity, PiLA identified 79.6\% of AS-positive participants and 66.7\% of AR-positive participants. These findings suggest that PiLA can effectively enrich individuals with clinically recognized AS and AR labels within the screened cohort, supporting its potential utility as a low-cost risk-enrichment tool for subsequent cardiovascular evaluation.

\FloatBarrier

\subsection{Hierarchical Comparison of Learning Paradigms}

To investigate the sources of performance improvement in PiLA, we performed a hierarchical comparison spanning supervised learning, generic large-scale pretraining paradigms, and physiology-guided pretraining strategies (Table~\ref{tab:unified_performance}).

\begin{table}[htbp]
  \centering
  \caption{Comparison of screening performance across supervised learning, generic self-supervised learning, and physiology-guided pretraining paradigms. The table presents the AUROC, Sensitivity at 70\% Specificity (S@70Sp), and Balanced Accuracy (BalAcc).}
  \label{tab:unified_performance}
  \begin{tabular}{lccc}
    \toprule
    Method & AUROC & S@70Sp & BalAcc \\
    \midrule

    \multicolumn{4}{l}{\textit{Panel A: AS }} \\
    \midrule

    \multicolumn{4}{l}{\textit{Supervised Baselines}} \\
    ResNet1D (Baseline) & 0.6951 & 0.5714 & 0.6545 \\
    TimesNet & 0.6956 & 0.5306 & 0.6539 \\

    \addlinespace
    \multicolumn{4}{l}{\textit{Generic SSL Paradigms}} \\
    SimCLR & 0.7554 & 0.6939 & 0.7135 \\
    Reconstruction & 0.7449 & 0.6531 & 0.7196 \\
    K-Means & 0.7487 & 0.7347 & 0.7335 \\

    \addlinespace
    \multicolumn{4}{l}{\textit{Physiology-guided Pretraining}} \\
    Physiological Feature Clustering & 0.7466 & 0.6939 & 0.7121 \\
    \textbf{PiLA (Ours)} & \textbf{0.8025} & \textbf{0.7959} & \textbf{0.7705} \\

    \midrule

    \multicolumn{4}{l}{\textit{Panel B: AR }} \\
    \midrule

    \multicolumn{4}{l}{\textit{Supervised Baselines}} \\
    ResNet1D (Baseline) & 0.6763 & 0.5476 & 0.6519 \\
    TimesNet & 0.6557 & 0.5208 & 0.6319 \\

    \addlinespace
    \multicolumn{4}{l}{\textit{Generic SSL Paradigms}} \\
    SimCLR & 0.6996 & 0.5476 & 0.6700 \\
    Reconstruction & 0.6866 & 0.5476 & 0.6843 \\
    K-Means & 0.7334 & 0.6429 & 0.6995 \\

    \addlinespace
    \multicolumn{4}{l}{\textit{Physiology-guided Pretraining}} \\
    Physiological Feature Clustering & 0.7083 & 0.5476 & 0.6833 \\
    \textbf{PiLA (Ours)} & \textbf{0.7669} & \textbf{0.6667} & \textbf{0.7343} \\

    \bottomrule
  \end{tabular}
\end{table}

We first evaluated two supervised learning baselines trained directly on the limited labeled dataset without large-scale pretraining, including ResNet1D and TimesNet~\cite{wu2023timesnet}. Both models achieved comparable performance, with AUROCs ranging from 0.65 to 0.70 across the AS and AR tasks. Although TimesNet has demonstrated strong performance across a variety of general time-series applications, it did not outperform the conventional ResNet1D baseline in the current AVD screening tasks. This observation suggests that, under limited-label conditions, increasing architectural complexity alone may be insufficient for learning robust representations of subtle AVD-related waveform alterations.

This bottleneck further highlights the challenge imposed by label scarcity. To leverage the large amount of available unlabeled PPG data, we next evaluated several generic self-supervised learning (SSL) paradigms that have been widely explored for addressing limited annotation in physiological time-series analysis~\cite{krishnan2022self, ding2024self,weng2025self}. Representative approaches included contrastive learning (SimCLR), signal reconstruction tasks, and clustering-based pseudo-supervision. Compared with purely supervised learning, these generic SSL approaches consistently improved downstream performance, highlighting the value of large-scale pretraining under label-limited conditions. Nevertheless, substantial performance gaps remained relative to PiLA across both AS and AR screening tasks.

For SimCLR, we implemented a standard augmentation pipeline including random crop with padding, Gaussian jitter, and magnitude scaling. We speculate that some perturbations introduced during augmentation may alter subtle waveform characteristics potentially relevant to AVD-related hemodynamic variation. Reconstruction objectives, meanwhile, may preferentially preserve global waveform contours rather than subtle local waveform dynamics. Similarly, direct K-Means clustering on raw signals likely captured dominant statistical variability unrelated to downstream cardiovascular physiology. Collectively, these findings suggest that while large-scale pretraining is beneficial, generic SSL objectives alone may be insufficient for robust extraction of disease-associated waveform representations from subtle physiological waveforms.

Given the limitations of generic SSL objectives, we further explored whether incorporating physiological knowledge into the pretraining task could improve representation learning. Indeed, integrating domain knowledge into learning—known as knowledge-guided machine learning—has become increasingly important for advancing AI in complex scientific domains~\cite{von2023informed}. While Physics-Informed Neural Networks (PINNs) have demonstrated success by embedding explicit physical laws~\cite{karniadakis2021physics}, such approaches typically rely on precise mathematical formulations and may be less adaptable to the heterogeneous physiological variability encountered in cardiovascular waveform analysis. Consequently, more flexible physiology-guided strategies based on signal-level feature points or rule-based supervision have recently attracted growing attention~\cite{liu2025physcl, maghsoodi2026domain, lee2021self}.

Drawing inspiration from this idea, we constructed a Physiological Feature Clustering pre-training strategy as a comparison baseline, utilizing six conventional pulse-wave metrics provided by the UK Biobank (including pulse rate and reflection index) to generate physiological pseudo-labels. Although this strategy improved performance relative to supervised learning, the gains remained limited compared with PiLA. These findings suggest that pretraining based primarily on physiological summary descriptors may be insufficient for learning disease-relevant waveform representations, potentially because substantial hemodynamic information is compressed or lost during low-dimensional feature abstraction.

In contrast, PiLA incorporates clinically motivated physiological waveform patterns (Tardus/WHP rules~\cite{meghraoui2024classifying}) into the construction of a large-scale proxy pretraining task. Rather than relying solely on generic signal transformations or summary-level physiological descriptors, the proposed framework leverages physiologically meaningful waveform characteristics to guide representation learning from nearly 170,000 unlabeled PPG recordings. Notably, PiLA consistently achieved the best overall performance across both AS and AR screening tasks. Taken together, these findings suggest that the effectiveness of large-scale physiological pretraining depends not only on access to unlabeled data, but also on the incorporation of physiologically relevant knowledge capable of guiding representation learning toward cardiovascular waveform characteristics associated with AVD-related waveform patterns.

\subsection{Architectural Ablation Analysis}

To better understand the contribution of individual design components in PiLA, we conducted an architectural ablation study focusing on two key design choices: (1) the physiologically structured three-stream representation of PPG, VPG, and APG, and (2) the dual-branch integration of pretrained and supervised learning. As summarized in Table~\ref{tab:ablation}, each architectural component contributed measurable performance improvements across the AS and AR screening tasks.

\begin{table}[htbp]
\centering
\caption{Architectural ablation analysis of PiLA.}
\label{tab:ablation}
\begin{tabular}{lcc}
\hline
\textbf{Model Variant} & \textbf{AUROC-AS} & \textbf{AUROC-AR} \\
\hline
Concat-Input Dual-Branch Model & 0.7373 & 0.7174 \\
Three-Stream Model without Co-Attention & 0.7780 & 0.7587 \\
Supervised-Only Three-Stream Model & 0.7371 & 0.6858 \\
Partially-Frozen Pretrained Model & 0.7412 & 0.7064 \\
\textbf{Full PiLA} & \textbf{0.8025} & \textbf{0.7669} \\
\hline
\end{tabular}
\end{table}

We first investigated whether PPG, VPG, and APG could be directly concatenated and processed using a conventional shared backbone. Although all three physiological signals were preserved as inputs, replacing dedicated stream-specific representation learning with direct input fusion resulted in a marked reduction in performance. In contrast, introducing separate streams for PPG, VPG, and APG substantially improved discrimination performance even in the absence of Co-Attention. These findings suggest that derivative-domain signals contain physiologically relevant information that is more effectively utilized through dedicated stream-wise representation learning rather than simple early fusion.

We next evaluated the contribution of the Co-Attention mechanism. Compared with direct stream fusion, incorporating Co-Attention further improved performance on both tasks, increasing AUROC from 0.7780 to 0.8025 for AS and from 0.7587 to 0.7669 for AR. This result suggests that explicit interaction between physiological streams provides additional benefit beyond independent stream modeling alone.

The branch-level ablations further supported the value of the dual-branch design. Removing either branch resulted in a consistent decline in performance relative to the full PiLA architecture. The Partially-Frozen Pretrained Three-Stream Model achieved performance comparable to, and slightly higher than, the Supervised-Only Three-Stream Model, suggesting that physiology-guided pretraining provides a useful initialization for downstream AVD screening. However, neither single-branch configuration matched the performance of the complete model. The full PiLA architecture consistently achieved the highest AUROC across both tasks, indicating that the integration of pretrained and supervised learning contributes to improved screening performance.

Taken together, these findings support the architectural rationale of PiLA. Effective AVD screening benefits from both physiologically structured multi-stream signal modeling and the integration of pretrained and supervised learning within a unified framework.

\subsection{Incremental Value of Handcrafted PPG Features}

To examine whether conventional handcrafted PPG features provided complementary information beyond the representations learned by PiLA, we designed a late-fusion experiment integrating handcrafted PPG features with deep representations learned by PiLA. Specifically, we extracted 22 conventional PPG morphological features spanning temporal intervals, waveform slopes, regional statistics, and waveform distribution descriptors, representing commonly used handcrafted pulse-wave characteristics.

Experimental results showed that introducing these handcrafted features led to a slight performance decline. The AUROC of the fused model decreased to 0.7415 for the AS task and 0.7514 for the AR task, both lower than the baseline performance achieved by the PiLA deep representation alone. These findings suggest that the learned deep representations already capture much of the discriminative waveform information relevant to downstream AVD screening tasks, while direct fusion with low-dimensional handcrafted features may introduce redundancy or additional noise.

One possible explanation lies in the limitations of handcrafted features when dealing with complex physiological confounding. Prior studies have shown that vascular aging and arterial stiffening can substantially overlap with waveform alterations associated with AS~\cite{hungerford2023interpretation}. Because handcrafted features are typically low-dimensional and manually predefined, they may have limited ability to represent complex non-linear waveform relationships under heterogeneous physiological conditions. In contrast, the representations learned by PiLA may preserve higher-order waveform dynamics embedded within derivative-domain signals, potentially contributing to improved robustness under physiologically heterogeneous settings.

\subsection{Clinical Confounding and Error Characterization}

Given the substantial demographic and clinical differences observed between AVD-positive participants and controls, we further evaluated the extent to which model performance could be influenced by measurable clinical confounding factors.

As a clinical-feature reference experiment, we trained a LightGBM classifier using routinely available demographic characteristics, vital signs, and comorbidity information. The clinical-feature baseline achieved AUROCs of 0.6999 for AS and 0.6605 for AR (Table~\ref{tab:confounding_analysis}), indicating that demographic characteristics and cardiovascular risk factors themselves carried substantial discriminative information for AVD screening. This observation is consistent with the established epidemiology of AVD, where age, sex, and cardiovascular disease burden are known risk factors and may also influence peripheral pulse-wave morphology.

\begin{table}[htbp]
\centering
\caption{Performance comparison under clinical-feature and covariate-balanced evaluation settings.}
\label{tab:confounding_analysis}
\begin{tabular}{lccc}
\toprule
\textbf{Model / Setting} & \textbf{Input} & \textbf{AS AUROC} & \textbf{AR AUROC} \\
\midrule
LightGBM Clinical Baseline & Clinical variables & 0.6999 & 0.6605 \\
PiLA (Covariate-balanced Cohort) & PPG waveform & 0.7066 & 0.6890 \\
PiLA (Original Cohort) & PPG waveform & 0.8025 & 0.7669 \\
\bottomrule
\end{tabular}
\end{table}

We further evaluated PiLA in a covariate-balanced downstream cohort constructed using propensity-score stratified sampling. Propensity-score balancing substantially reduced baseline differences across most demographic and clinical characteristics.

In this more stringent covariate-balanced setting, PiLA achieved AUROCs of 0.7066 for AS and 0.6890 for AR, compared with 0.8025 and 0.7669 in the original cohort, respectively (Table~\ref{tab:confounding_analysis}). The reduction in performance following propensity-score balancing was expected and suggests that demographic characteristics and cardiovascular comorbidities contributed to part of the discrimination observed in the original cohort. However, despite substantial balancing of most demographic characteristics and clinical risk factors, PiLA maintained moderate discrimination performance in the covariate-balanced cohort. These findings suggest that while clinical confounding accounts for a portion of the observed screening performance, PPG waveform morphology itself retains information associated with AVD risk beyond the matched clinical characteristics.

To further characterize how clinical factors manifested in individual model predictions, we performed an error analysis in the independent test cohort (Table~\ref{tab:error_analysis}). No significant differences were observed between TP and FN cases across age, BMI, arterial stiffness, hypertension, diabetes mellitus, CAD/MI, or atrial fibrillation for either AS or AR. These findings suggest that missed positive cases were not concentrated within any obvious demographic or clinical subgroup, providing little evidence of major systematic clinical bias among incorrectly classified positive participants.

\begin{table*}[htbp]
\centering
\caption{
Error characterization analysis in the independent test cohort. Participants were stratified according to model-derived risk ranking and disease labels. Continuous variables are presented as mean $\pm$ SD and categorical variables as percentages. P1 indicates comparisons between TP and FN groups, whereas P2 indicates comparisons between FP and TN groups.
}
\label{tab:error_analysis}

\footnotesize
\setlength{\tabcolsep}{3pt}

\begin{tabular*}{\textwidth}{@{\extracolsep{\fill}}lcccccc}
\toprule
\textbf{Variable} &
\textbf{TP} &
\textbf{FN} &
\textbf{TN} &
\textbf{FP} &
\textbf{P1} &
\textbf{P2} \\
\midrule

\multicolumn{7}{l}{\textbf{Panel A: AS}}\\
\midrule

Age (years)
& 62.0 $\pm$ 8.1
& 61.4 $\pm$ 7.1
& 56.3 $\pm$ 8.1
& 59.5 $\pm$ 6.7
& 0.790
& $<$0.001 \\

Male sex (\%)
& 64.7
& 62.5
& 45.8
& 25.8
& 1.000
& $<$0.001 \\

BMI (kg/m$^2$)
& 29.6 $\pm$ 3.7
& 30.1 $\pm$ 5.5
& 27.4 $\pm$ 4.8
& 28.9 $\pm$ 5.8
& 0.706
& 0.012 \\

Arterial Stiffness Index
& 9.3 $\pm$ 3.8
& 9.7 $\pm$ 3.0
& 9.1 $\pm$ 3.0
& 10.5 $\pm$ 3.5
& 0.659
& $<$0.001 \\

Hypertension (\%)
& 88.2
& 87.5
& 37.4
& 60.2
& 1.000
& $<$0.001 \\

Diabetes Mellitus (\%)
& 23.5
& 31.2
& 10.9
& 16.1
& 0.743
& 0.169 \\

CAD / MI (\%)
& 76.5
& 71.9
& 10.7
& 17.2
& 1.000
& 0.084 \\

Atrial Fibrillation (\%)
& 52.9
& 40.6
& 7.3
& 11.8
& 0.548
& 0.148 \\

\midrule

\multicolumn{7}{l}{\textbf{Panel B: AR}}\\
\midrule

Age (years)
& 61.2 $\pm$ 9.4
& 62.2 $\pm$ 5.2
& 56.5 $\pm$ 8.1
& 58.2 $\pm$ 7.6
& 0.699
& 0.034 \\

Male sex (\%)
& 68.8
& 65.4
& 45.6
& 27.7
& 1.000
& $<$0.001 \\

BMI (kg/m$^2$)
& 29.5 $\pm$ 3.9
& 27.1 $\pm$ 5.9
& 27.5 $\pm$ 4.9
& 28.6 $\pm$ 5.7
& 0.127
& 0.058 \\

Arterial Stiffness Index
& 9.1 $\pm$ 4.1
& 9.8 $\pm$ 3.7
& 9.2 $\pm$ 3.0
& 9.5 $\pm$ 3.2
& 0.576
& 0.364 \\

Hypertension (\%)
& 81.2
& 73.1
& 38.6
& 56.4
& 0.715
& $<$0.001 \\

Diabetes Mellitus (\%)
& 37.5
& 23.1
& 11.3
& 13.8
& 0.483
& 0.497 \\

CAD / MI (\%)
& 81.2
& 57.7
& 11.4
& 18.1
& 0.180
& 0.067 \\

Atrial Fibrillation (\%)
& 68.8
& 50.0
& 7.1
& 10.6
& 0.338
& 0.215 \\

\bottomrule
\end{tabular*}
\end{table*}

In contrast, individuals assigned to the high-risk group despite lacking AS/AR labels exhibited a greater burden of established cardiovascular risk factors. For AS, FP cases were older and demonstrated higher BMI, arterial stiffness, and hypertension prevalence than TN cases. Similar trends were observed for AR, where FP cases were older and more frequently hypertensive. Interestingly, FP cases contained a lower proportion of male participants than TN cases despite the generally higher prevalence of both AS and AR among men reported in epidemiological studies. This observation suggests that sex alone is unlikely to account for the model's risk assignments and further supports the multifactorial nature of the learned representations. Together with the clinical-feature baseline and covariate-balanced cohort analysis, these observations suggest that conventional clinical risk factors contribute to model predictions but do not fully explain model behavior. Rather, PiLA appears to integrate both clinical-risk-related and waveform-specific information when generating predictions.

\subsection{AS--AR Subtype Discrimination Analysis}

To directly evaluate whether PiLA captured subtype-specific information beyond general AVD-related signals, we performed an AS--AR subtype discrimination analysis restricted to participants with isolated AS or isolated AR. Controls and mixed AVD cases were excluded to avoid confounding subtype discrimination with disease-versus-control separation or ambiguous mixed-valve phenotypes.

As shown in Table~\ref{tab:asar_discrimination}, the score-based analysis using the difference between the AS and AR prediction scores achieved an AUROC of 0.682. To further assess whether subtype-discriminative information was present in the learned representations themselves, we trained an L2-regularized logistic regression linear probe on frozen PiLA embeddings. The linear probe achieved a comparable AUROC of 0.680.

\begin{table}[htbp]
\centering
\caption{AS--AR subtype discrimination analysis among isolated AS and isolated AR cases.}
\label{tab:asar_discrimination}
\begin{tabular}{lc}
\toprule
Method & AUROC \\
\midrule
Score-based discrimination & 0.682 \\
Representation-based linear probe & 0.680 \\
\bottomrule
\end{tabular}
\end{table}

The similar performance of the score-based and representation-based analyses suggests that subtype-discriminative information is already encoded within the learned representations rather than arising solely from the task-specific output heads. These findings indicate that PiLA captures information that partially differentiates AS from AR, although the moderate level of discrimination also suggests substantial overlap between the two conditions in terms of PPG morphology. Collectively, the results support the presence of subtype-related information within the learned representations while reinforcing the intended role of PiLA as a screening and risk-enrichment framework rather than a definitive AS/AR subtype classifier.

\subsection{Temporal Sensitivity Across Diagnostic Horizons}

\begin{figure*}[htbp]
\centering

\begin{subfigure}[b]{0.48\textwidth}
    \centering
    \includegraphics[width=\linewidth]{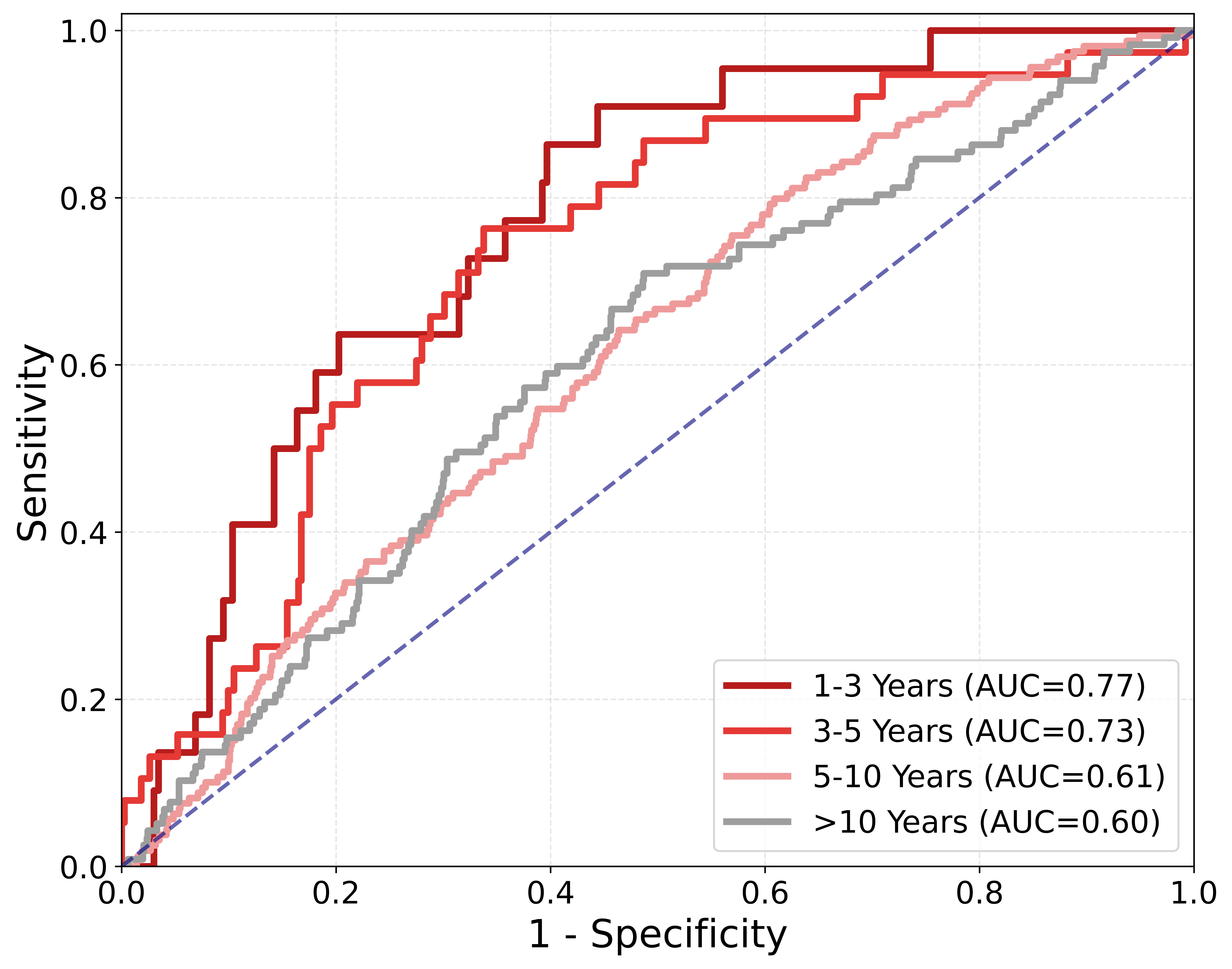}
    \caption{}
    \label{fig:temporal_roc_as}
\end{subfigure}
\hfill
\begin{subfigure}[b]{0.48\textwidth}
    \centering
    \includegraphics[width=\linewidth]{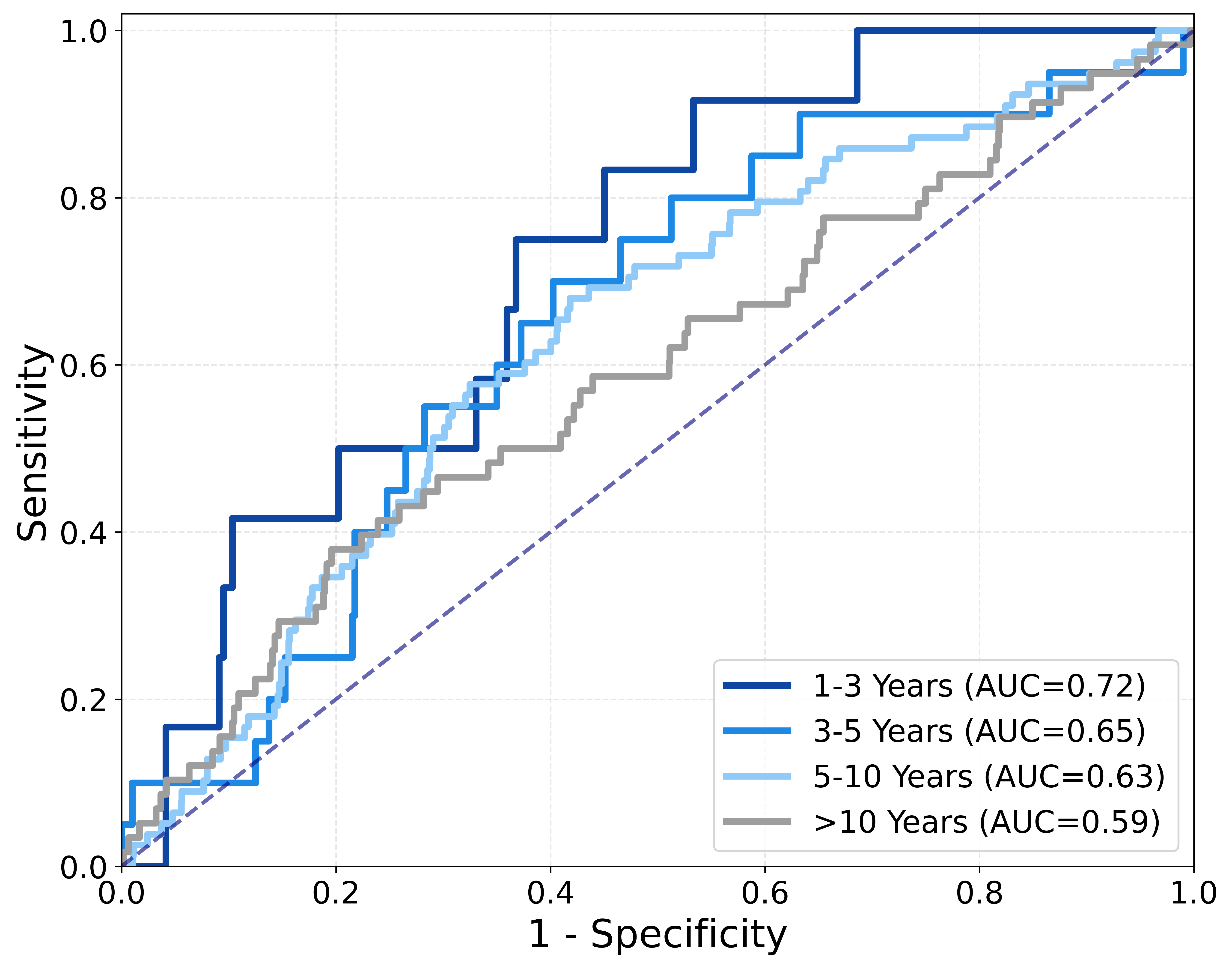}
    \caption{}
    \label{fig:temporal_roc_ar}
\end{subfigure}

\vspace{0.5em}

\begin{subfigure}[b]{0.48\textwidth}
    \centering
    \includegraphics[width=\linewidth]{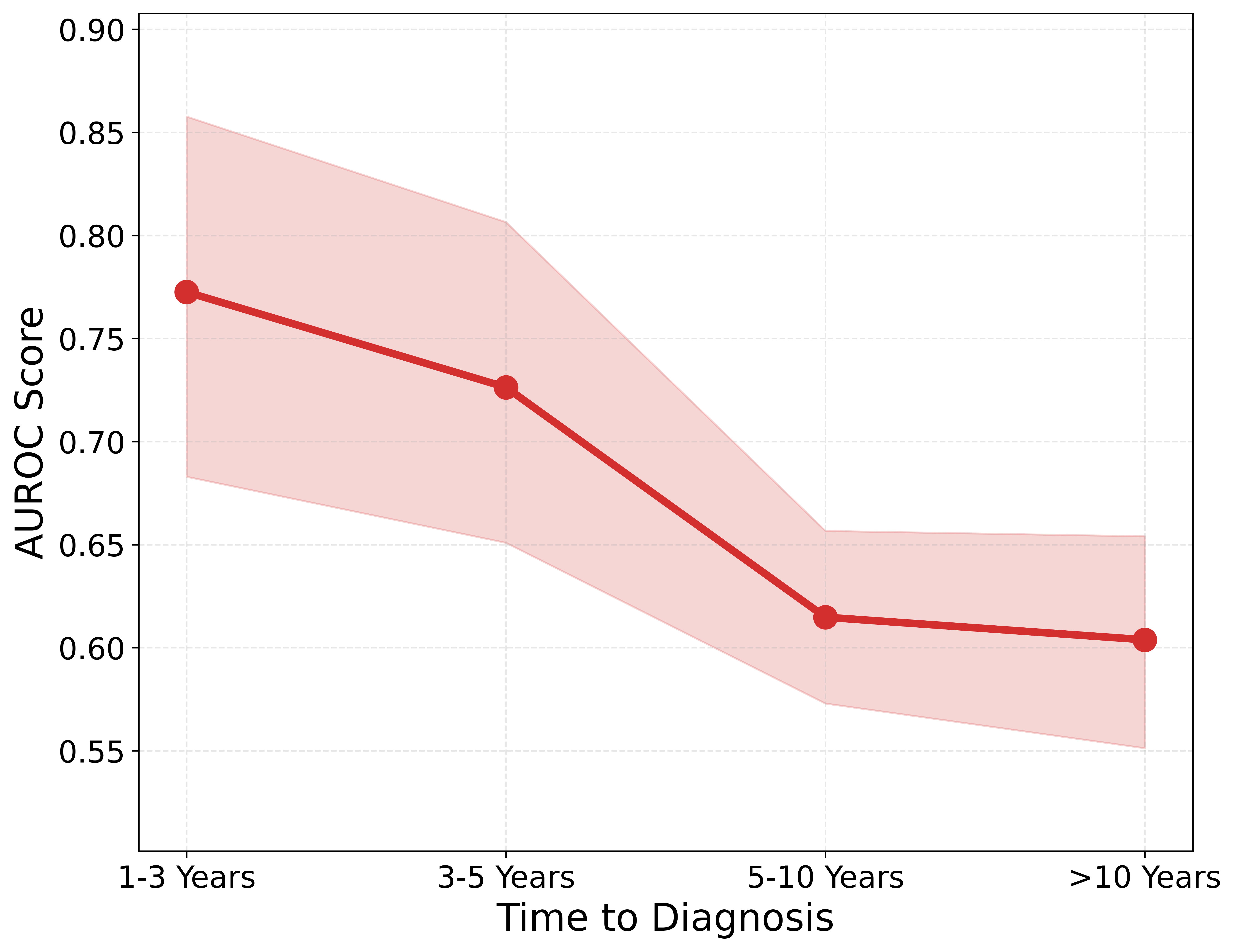}
    \caption{}
    \label{fig:temporal_trend_as}
\end{subfigure}
\hfill
\begin{subfigure}[b]{0.48\textwidth}
    \centering
    \includegraphics[width=\linewidth]{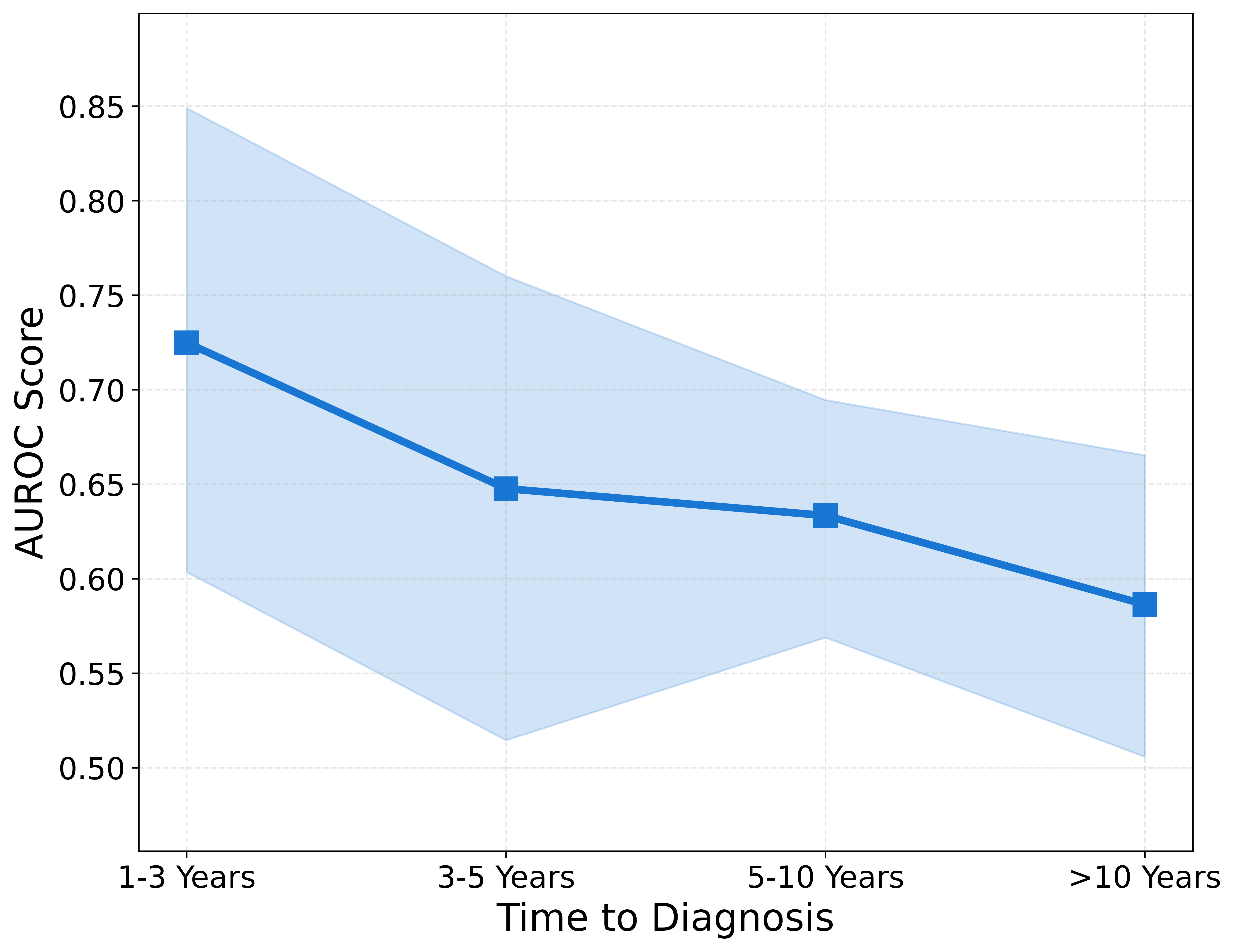}
    \caption{}
    \label{fig:temporal_trend_ar}
\end{subfigure}

\caption{
Temporal sensitivity analysis of PiLA across different diagnostic horizons. a) ROC curves for AS stratified by time-to-diagnosis intervals. b) ROC curves for AR stratified by time-to-diagnosis intervals. c) AUROC trend for AS across diagnostic horizons. d) AUROC trend for AR across diagnostic horizons. Shaded areas represent 95\% confidence intervals. For both AS and AR, screening performance was highest among participants diagnosed closest to the time of PPG acquisition and generally declined as the diagnostic horizon increased.
}
\label{fig:temporal_sensitivity}
\end{figure*}

To evaluate the temporal relationship between model predictions and subsequent clinical recognition of AVD, we stratified participants according to the interval between PPG acquisition and the first recorded diagnosis of AS or AR. The quantitative results are summarized in Figure~\ref{fig:temporal_sensitivity} and Table~\ref{tab:temporal_performance}.

\begin{table}[htbp]
  \caption{Model Performance Stratified by Time-to-Diagnosis Intervals}
  \label{tab:temporal_performance}
  \begin{tabular}{llcc}
    \toprule
    \textbf{Disease} & \textbf{Time Interval} & \textbf{AUROC (95\% CI)} & \textbf{Sens @ 70\% Spec} \\
    \midrule
    \multirow{4}{*}{AS}
    & 1--3 Years & 0.773 (0.681--0.858) & 63.6\% \\
    & 3--5 Years & 0.726 (0.649--0.806) & 65.8\% \\
    & 5--10 Years & 0.615 (0.572--0.656) & 43.4\% \\
    & $>$10 Years & 0.604 (0.552--0.654) & 46.2\% \\
    \midrule
    \multirow{4}{*}{AR}
    & 1--3 Years & 0.725 (0.602--0.843) & 50.0\% \\
    & 3--5 Years & 0.648 (0.523--0.762) & 55.0\% \\
    & 5--10 Years & 0.633 (0.565--0.695) & 51.3\% \\
    & $>$10 Years & 0.586 (0.505--0.665) & 46.6\% \\
    \bottomrule
  \end{tabular}
\end{table}

For both AS and AR, discrimination performance was highest among participants diagnosed within 1--3 years after PPG acquisition and showed a tendency toward lower performance at more distant diagnostic horizons. For AS, AUROC decreased from 0.773 in the 1--3 year interval to 0.604 in the $>$10 year interval. Similarly, for AR, AUROC decreased from 0.725 to 0.586 across the same diagnostic horizons.

Although discrimination performance gradually decreased at longer diagnostic horizons, performance remained above random discrimination across all temporal strata. Notably, even among participants who were not clinically recognized until more than 10 years after PPG acquisition, the model maintained AUROC values of 0.604 for AS and 0.586 for AR.

These observations are consistent with the possibility that disease-related cardiovascular alterations become increasingly reflected in peripheral pulse waveforms as clinical recognition approaches. Collectively, the findings suggest that PiLA captures disease-related physiological information that may already be present years before future clinical recognition of AVD. The persistence of above-random discrimination across extended diagnostic horizons further supports the potential utility of PPG-based representation learning for opportunistic screening and risk enrichment in asymptomatic populations.

\subsection{Long-term Risk Stratification and Independent Prognostic Value}

To further evaluate the longitudinal risk stratification capability of PiLA beyond conventional clinical variables, we performed multivariable Cox regression and propensity score matching analyses.

In the multivariable Cox regression analysis of the incident follow-up cohort, we constructed three progressive adjustment models to quantify the association between PiLA scores and the risk of future aortic valve disease. Model 1 adjusted only for basic demographic characteristics including age, sex, and BMI. Model 2 further adjusted for systolic blood pressure, diastolic blood pressure, smoking status, and arterial stiffness index to account for hemodynamic, vascular, and lifestyle-related factors. Model 3 additionally adjusted for hypertension, diabetes, hyperlipidemia, chronic kidney disease, coronary artery disease/myocardial infarction, atrial fibrillation, and stroke, providing the most comprehensively adjusted assessment of independence.

As shown in Table~\ref{tab:multivariable_cox}, model-predicted risk scores were significantly associated with future clinically recognized AVD events across all adjustment models. In Model 1, each one-standard-deviation increase in the PiLA score was associated with a hazard ratio of 1.469 (95\% CI: 1.397--1.545, $p = 1.04 \times 10^{-50}$) for AS and 1.352 (95\% CI: 1.301--1.404, $p = 3.67 \times 10^{-54}$) for AR.
Notably, even after progressively adjusting for additional hemodynamic variables, lifestyle factors, and major cardiovascular comorbidities in Models 2 and 3, the observed associations remained stable with only modest attenuation of the hazard ratios (HR). In the fully adjusted Model 3, the hazard ratios remained 1.405 (95\% CI: 1.334--1.480, $p = 6.55 \times 10^{-38}$) for AS and 1.318 (95\% CI: 1.268--1.370, $p = 3.94 \times 10^{-44}$) for AR, both retaining strong statistical significance. This consistency across adjustment models suggests that the association between model-predicted risk scores and future clinically recognized AVD events is not fully explained by the measured clinical risk factors included in the adjustment models.

\begin{table}[htbp]
  \caption{Multivariable Cox Regression Analysis for Incident AVD in the Incident Follow-up Cohort}
  \label{tab:multivariable_cox}
  \centering
  \footnotesize
  \setlength{\tabcolsep}{3pt}
  \begin{tabularx}{\linewidth}{
    @{}
    >{\raggedright\arraybackslash}p{0.20\linewidth}
    >{\centering\arraybackslash}p{0.06\linewidth}
    >{\raggedright\arraybackslash}X
    >{\centering\arraybackslash}p{0.20\linewidth}
    >{\centering\arraybackslash}p{0.14\linewidth}
    @{}
  }
    \toprule
    \textbf{Outcome} & \textbf{Model} & \textbf{Adjustments} & \textbf{HR (95\% CI)} & \textbf{\textit{P}-value} \\
    \midrule
    \multirow{3}{=}{\textbf{AS}}
    & 1 & Age, Sex, BMI & 1.469 (1.397--1.545) & $1.04 \times 10^{-50}$ \\
    & 2 & Model 1 + SBP, DBP, Smoking, ASI & 1.453 (1.380--1.530) & $8.07 \times 10^{-46}$ \\
    & 3 & Model 2 + Hypertension, Diabetes, Hyperlipidemia, CKD, CAD/MI, AF, Stroke & 1.405 (1.334--1.480) & $6.55 \times 10^{-38}$ \\
    \midrule
    \multirow{3}{=}{\textbf{AR}}
    & 1 & Age, Sex, BMI & 1.352 (1.301--1.404) & $3.67 \times 10^{-54}$ \\
    & 2 & Model 1 + SBP, DBP, Smoking, ASI & 1.350 (1.299--1.403) & $3.34 \times 10^{-52}$ \\
    & 3 & Model 2 + Hypertension, Diabetes, Hyperlipidemia, CKD, CAD/MI, AF, Stroke & 1.318 (1.268--1.370) & $3.94 \times 10^{-44}$ \\
    \bottomrule
  \end{tabularx}
\end{table}

To further validate this finding under more controlled conditions, we conducted a survival analysis in a propensity-score balanced cohort. We precisely matched each incident case with control individuals using propensity scores estimated from demographic characteristics, arterial stiffness, and major cardiovascular comorbidities, thereby ensuring a Standardized Mean Difference (SMD) below 0.1 across all matched covariates. This design reduced measured baseline differences in traditional clinical risk factors and enabled evaluation of waveform-based risk stratification under a more balanced physiological setting.

\begin{figure*}[htbp]
  \centering

  \begin{subfigure}[b]{0.48\textwidth}
    \centering
    \includegraphics[width=\linewidth]{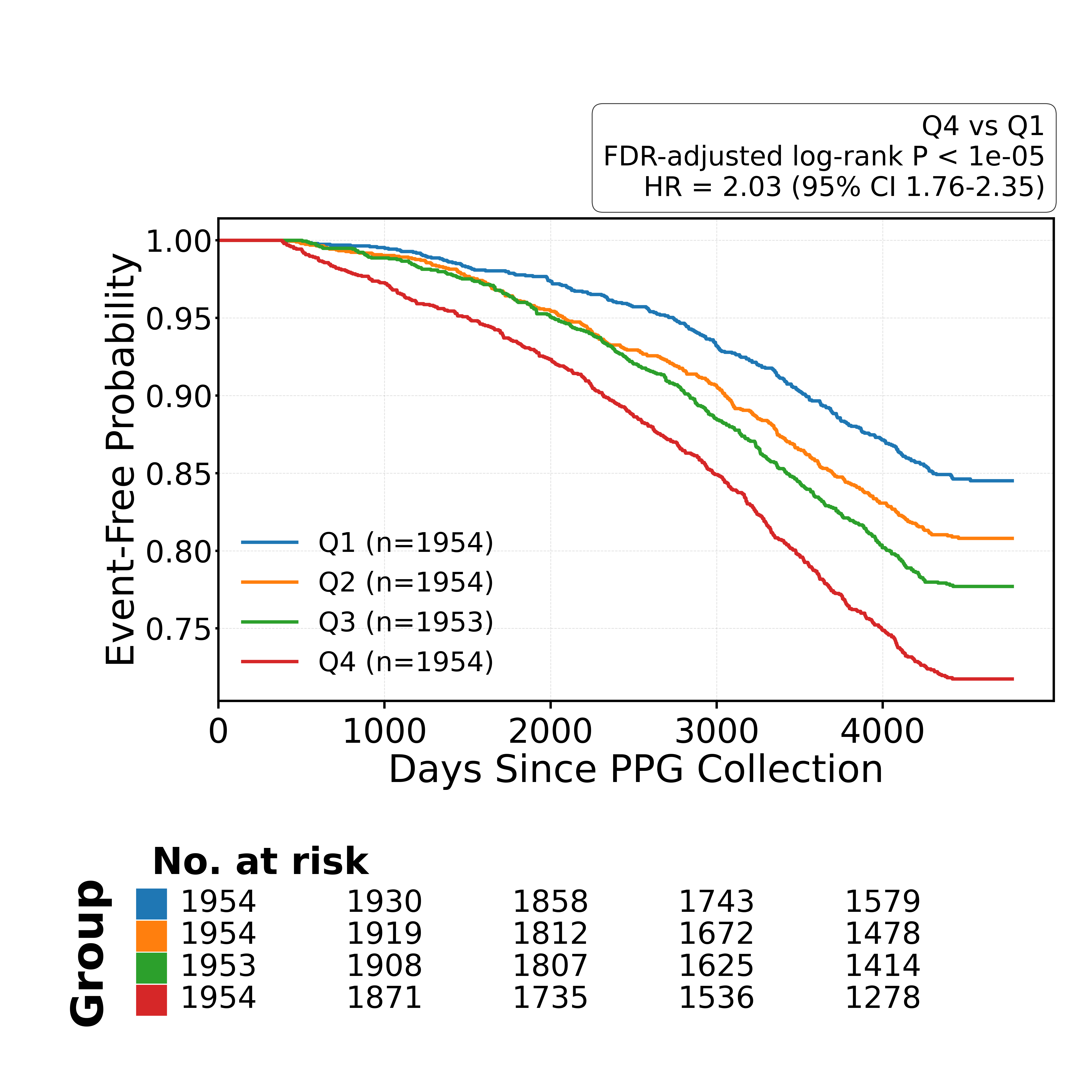}
    \caption{}
    \label{fig:as_quartiles}
  \end{subfigure}
  \hfill
  \begin{subfigure}[b]{0.48\textwidth}
    \centering
    \includegraphics[width=\linewidth]{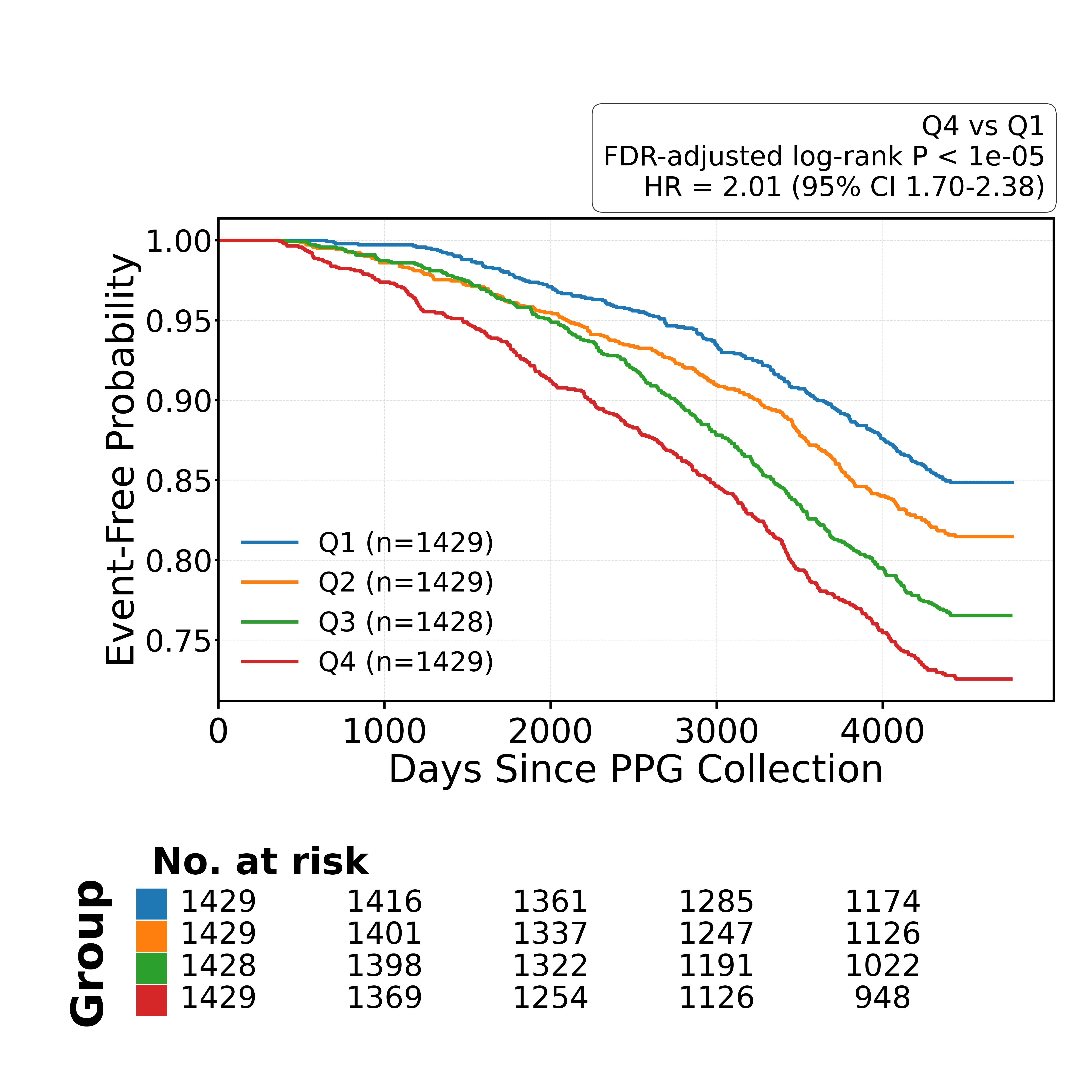}
    \caption{}
    \label{fig:ar_quartiles}
  \end{subfigure}

  \vspace{0.3cm}

  \begin{subfigure}[b]{0.48\textwidth}
    \centering
    \includegraphics[width=\linewidth]{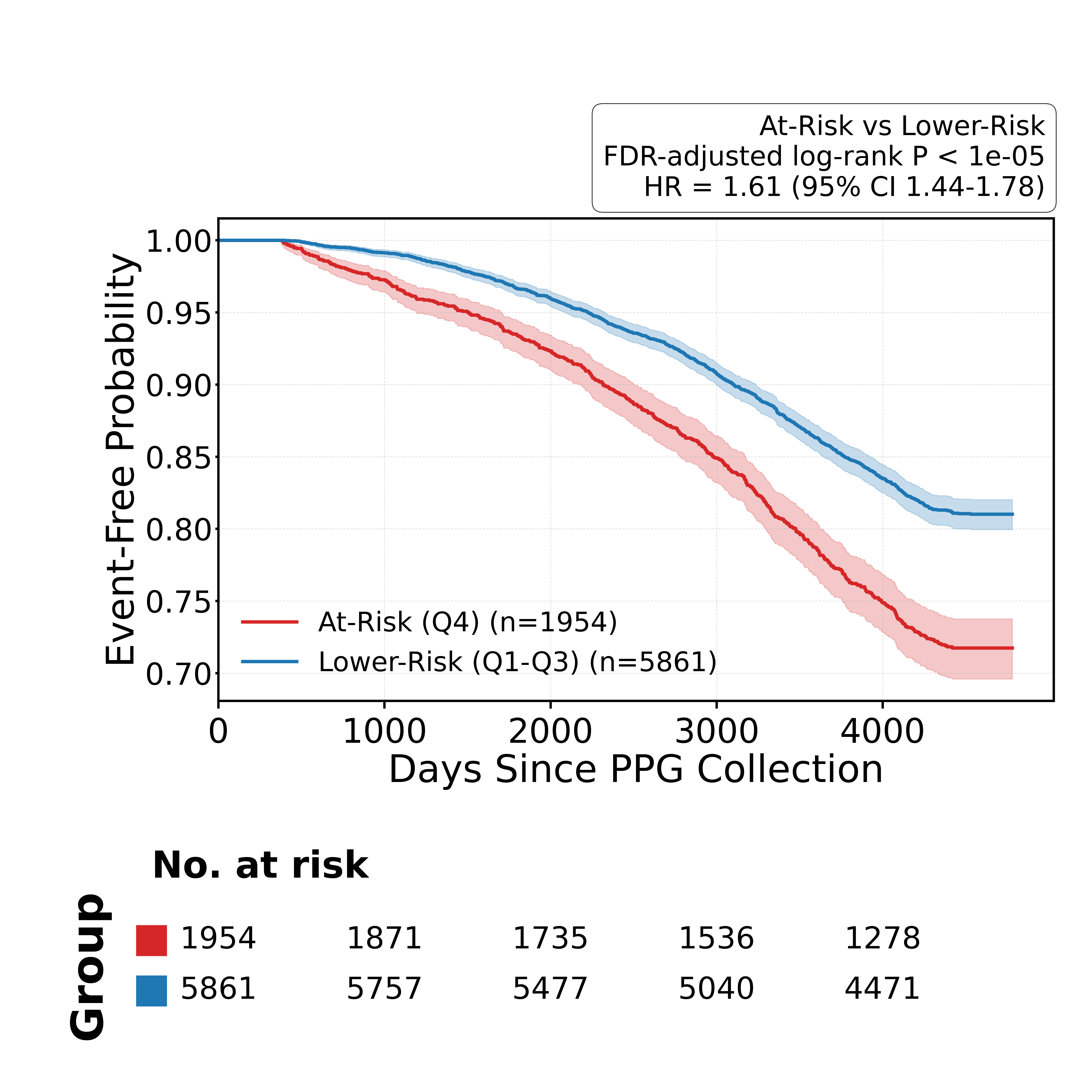}
    \caption{}
    \label{fig:as_binary}
  \end{subfigure}
  \hfill
  \begin{subfigure}[b]{0.48\textwidth}
    \centering
    \includegraphics[width=\linewidth]{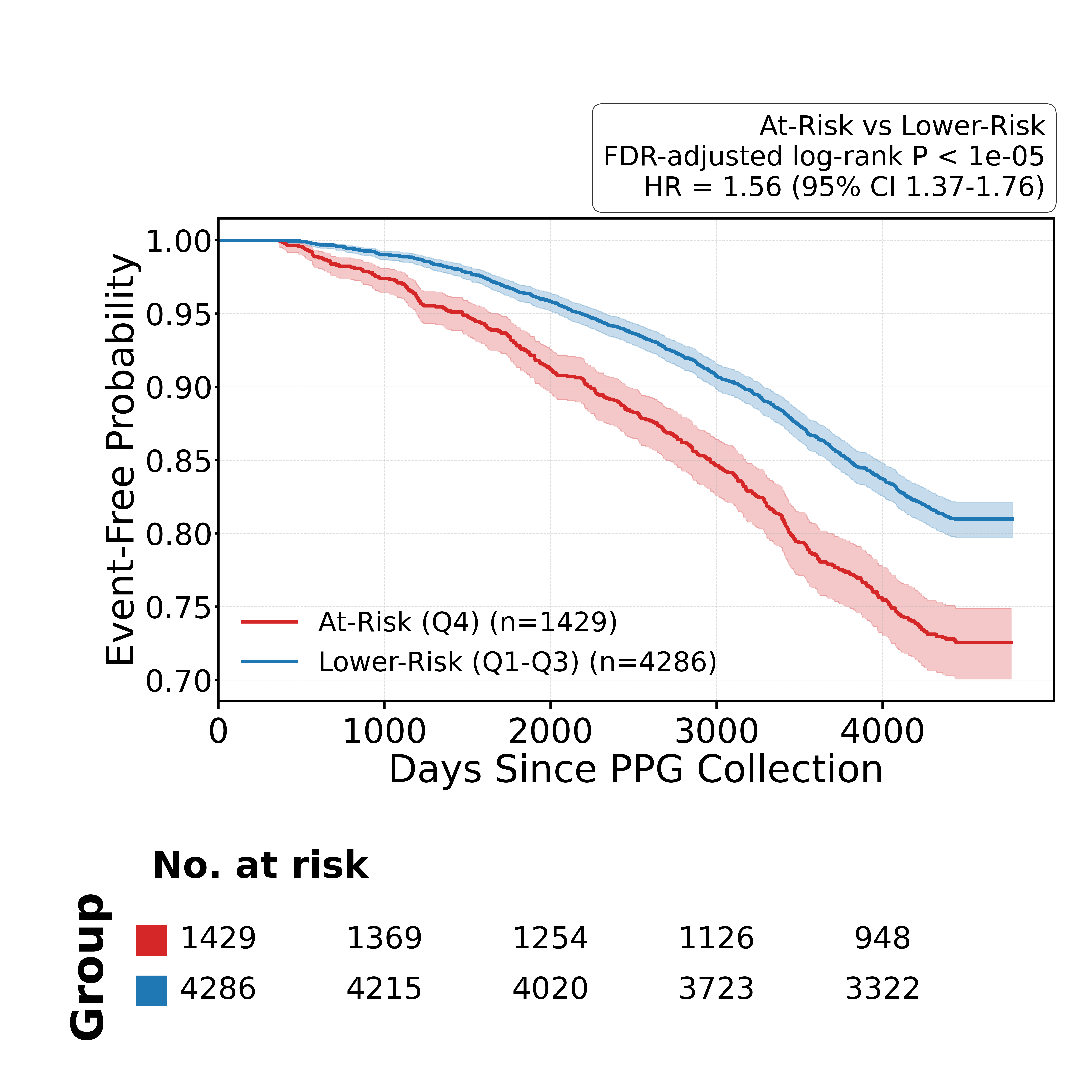}
    \caption{}
    \label{fig:ar_binary}
  \end{subfigure}

  \caption{Kaplan--Meier survival analysis in the propensity-score matched cohort. a) Event-free survival curves stratified by quartiles (Q1--Q4) of the predicted AS risk score. b) Event-free survival curves stratified by quartiles (Q1--Q4) of the predicted AR risk score. c) Event-free survival curves comparing participants in the highest predicted AS risk quartile (Q4; top 25\%) with all remaining participants (Q1--Q3). d) Event-free survival curves comparing participants in the highest predicted AR risk quartile (Q4; top 25\%) with all remaining participants (Q1--Q3). Hazard ratios compare the highest-risk group with the reference group in each analysis. Shaded areas represent 95\% confidence intervals, and risk tables are shown below each plot. Event-free survival progressively decreased with increasing predicted risk for both AS and AR.}
  \label{fig:survival_analysis}
\end{figure*}

In this highly balanced cohort (AS group $n=7,815$; AR group $n=5,715$), Kaplan-Meier event-free survival curves revealed a significant and monotonic dose-response gradient between model-predicted probabilities and long-term incidence risk. Events occurring within the first year after PPG acquisition were excluded to reduce potential contamination from clinically unrecognized prevalent disease at baseline. Following this landmark period, for AS (Figure~\ref{fig:survival_analysis}a), the survival trajectories of the four risk quartiles (Q1--Q4) showed clear stepwise separation. Cox proportional hazards regression analysis indicated that individuals classified into the highest risk group (Q4) had a long-term incidence risk 2.03 times that of the lowest risk group (Q1) (95\% CI: 1.76--2.35, $p<0.0001$). This finding suggests that model-derived waveform representations remained associated with future clinically recognized AS events even after extensive balancing of measured physiological and clinical characteristics. For AR, which involves more complex pathological mechanisms (Figure~\ref{fig:survival_analysis}b), the model also demonstrated robust stratification capability, with a Hazard Ratio of 2.01 for the Q4 group relative to Q1 (95\% CI: 1.70--2.38, $p<0.0001$), supporting the presence of a longitudinal association between model-derived risk scores and future clinically recognized AR events.

Further examining the effectiveness of clinical screening strategies, we defined the top 25\% of risk scores as ``High-Risk Screening Targets.'' As shown in Figure~\ref{fig:survival_analysis}c and d, their survival curves significantly deviated from the remaining 75\% low-risk population early in the follow-up period (Log-Rank $p<0.0001$). Individuals flagged as high-risk by the model had a 61\% increased risk of developing AS (HR=1.61) and a 56\% increased risk for AR (HR=1.56). Collectively, these findings suggest that the waveform representations learned by PiLA retain longitudinal risk-associated information beyond traditional cardiovascular risk factors, supporting their potential utility for risk-enrichment in large-scale physiological screening settings.

\FloatBarrier

\subsection{Subgroup Analysis and Physiological Heterogeneity}
\label{sec:subgroup_analysis}

\begin{figure}[htbp]
  \centering

  \begin{subfigure}{\textwidth}
    \centering
    \includegraphics[width=0.85\textwidth]{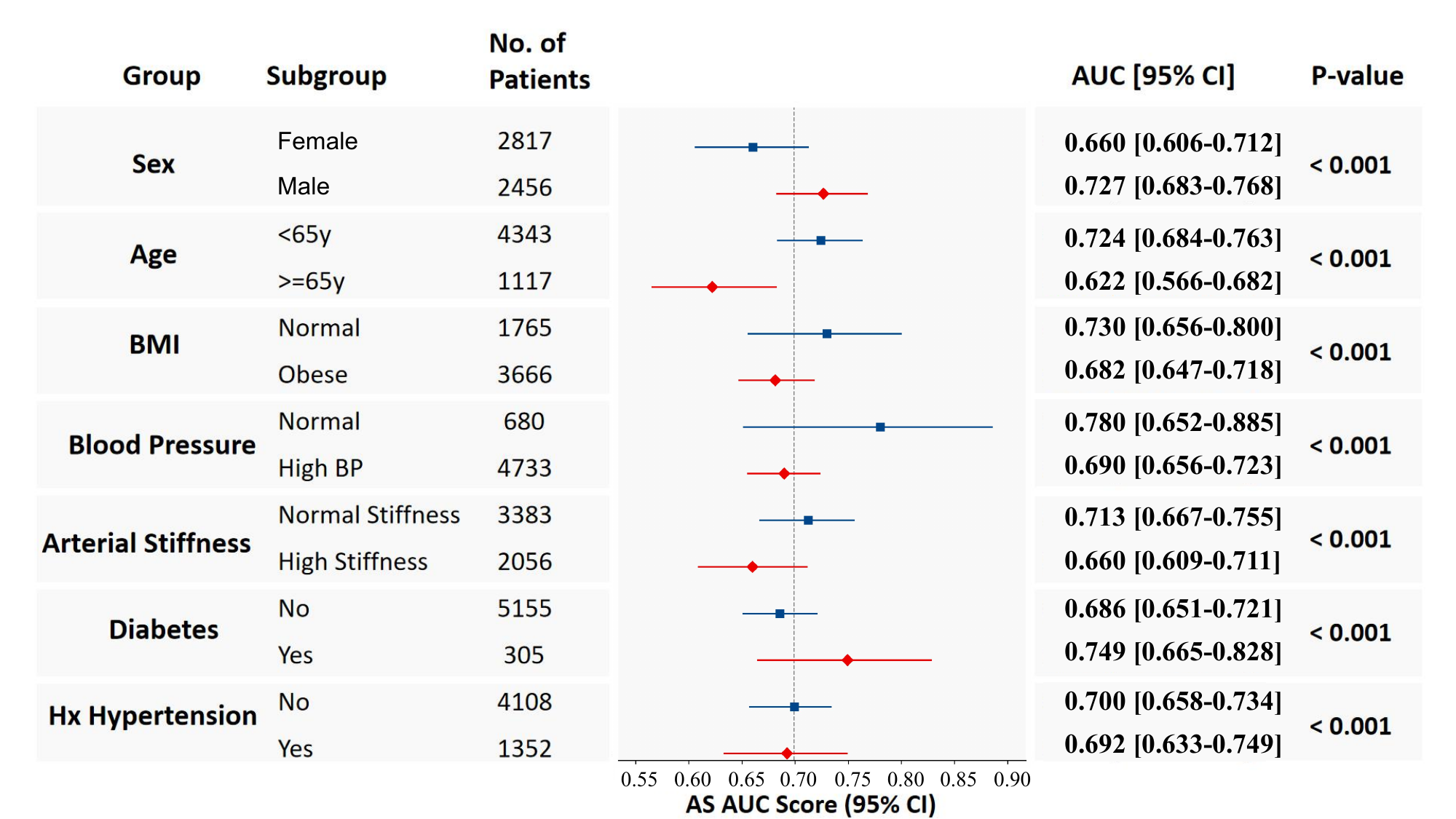}
    \caption{}
  \end{subfigure}

  \vspace{0.3em}

  \begin{subfigure}{\textwidth}
    \centering
    \includegraphics[width=0.85\textwidth]{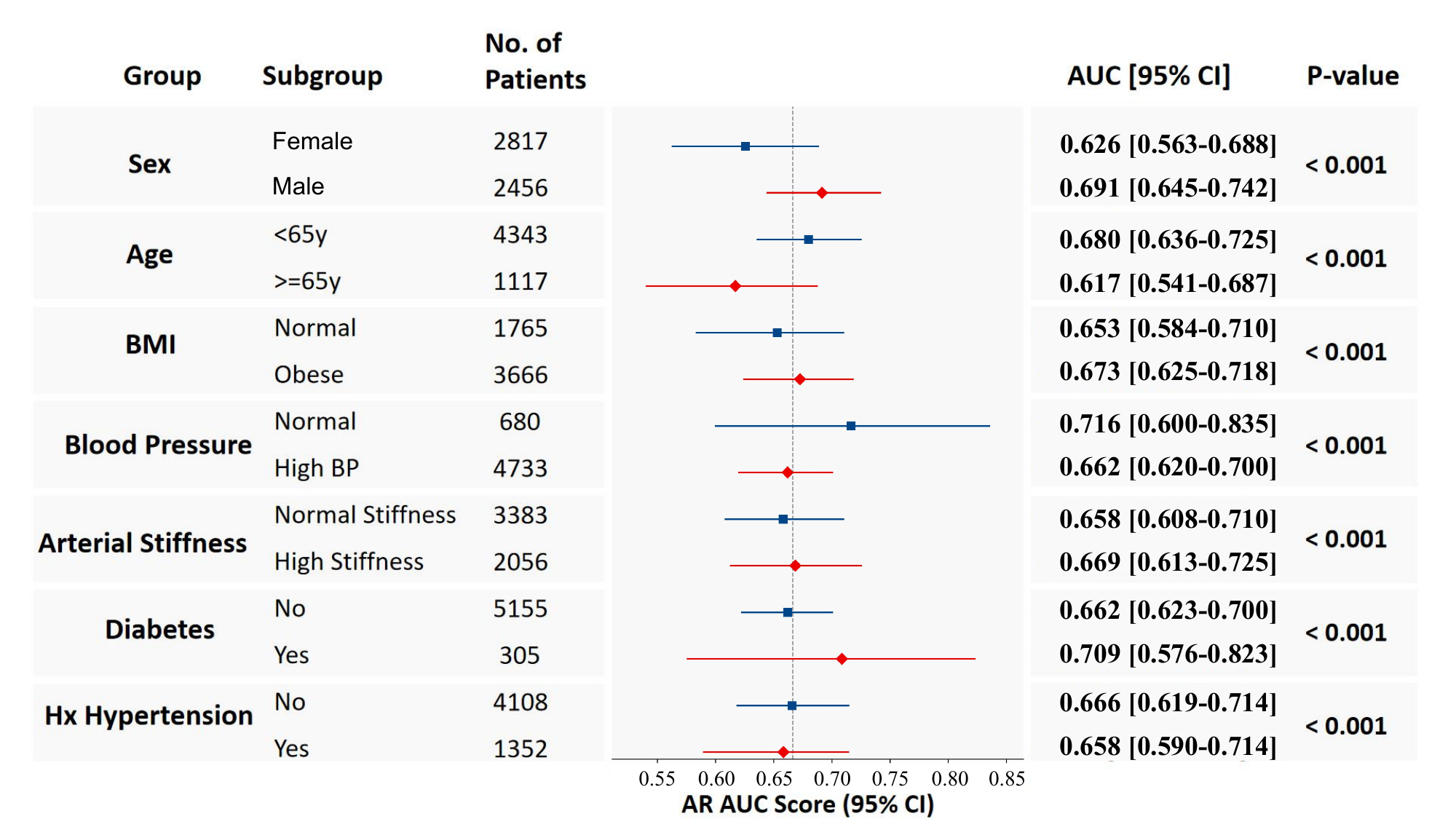}
    \caption{}
  \end{subfigure}

  \caption{
Forest plots of subgroup analyses across major physiological backgrounds. a) Subgroup analysis for AS screening. b) Subgroup analysis for AR screening. Points indicate AUROC estimates for each subgroup, and horizontal lines represent 95\% confidence intervals. Model performance varied across physiological subgroups, indicating heterogeneity in screening performance under different physiological backgrounds.
}
  \label{fig:forest_plots}
\end{figure}

To assess the consistency of model performance across diverse physiological backgrounds and evaluate the physiological relevance of the learned waveform representations, we conducted a comprehensive subgroup analysis across the 5,460 subjects included in the downstream cohort. The forest plots (Figure~\ref{fig:forest_plots}) show that while the model maintains discriminative ability across all subgroups, performance varied across physiological backgrounds, suggesting that physiological state may influence the manifestation of disease-related waveform characteristics.

First, factors related to vascular compliance and arterial remodeling appeared to influence performance most clearly in the AS screening task. Participants with lower arterial stiffness achieved higher discrimination performance than those with elevated stiffness (AUROC 0.713 vs. 0.660), while a similar pattern was observed across age strata, with higher performance in participants younger than 65 years (AUROC 0.724 vs. 0.622, $P<0.001$). Performance was also modestly lower among obese participants than among those with normal BMI. Obesity is closely associated with accelerated vascular aging and increased arterial stiffness, and a broadly similar performance pattern was observed in this subgroup. The consistency of these observations across age, arterial stiffness, and BMI strata is notable. AS-related waveform characteristics arise from alterations in ventricular outflow and pulse-wave transmission, whereas vascular aging, arterial stiffening, and obesity-related vascular remodeling can independently alter peripheral pulse-wave morphology and therefore influence the expression of disease-related waveform signatures. The reduced performance observed in populations with greater vascular burden therefore suggests that background vascular properties may partially obscure or alter the manifestation of AS-related hemodynamic signatures in peripheral PPG signals. Conversely, the superior performance observed in participants with better vascular compliance is broadly consistent with the interpretation that the model relies on physiologically meaningful waveform characteristics associated with AS rather than solely on demographic information.

Second, physiological loading conditions appeared to have a stronger influence on AR screening performance. The most apparent subgroup difference was observed across blood pressure strata, where the model achieved an AUROC of 0.716 in normotensive participants compared with 0.662 in hypertensive participants. Because AR is fundamentally characterized by abnormal diastolic flow dynamics, this pattern suggests that background hemodynamic conditions may influence the expression of AR-related waveform characteristics in peripheral pulse signals. A degree of sexual dimorphism was also observed, particularly in the AS task, where performance was consistently higher among male participants. Prior studies have reported sex-specific differences in AS pathophysiology and cardiac remodeling patterns, with males more frequently exhibiting pronounced valvular calcification and ventricular remodeling, whereas females often present with lower-flow phenotypes and greater myocardial fibrosis~\cite{saeed2020sex}. These biological differences may contribute to variation in how disease-related hemodynamic alterations are reflected in peripheral pulse waveforms.

In summary, the subgroup analysis highlights the substantial physiological heterogeneity encountered in real-world populations. Given the extreme scarcity of positive samples in this study, the model cannot fully cover the waveform variation space of all complex physiological subtypes. Nevertheless, several performance trends were consistently observed across related physiological dimensions, particularly those associated with vascular compliance and hemodynamic loading. These patterns were broadly aligned with established cardiovascular physiology, supporting the physiological relevance of the waveform representations learned by PiLA and highlighting the influence of physiological background on PPG-based AVD screening performance.

\FloatBarrier

\subsection{Physiological Attention Pattern Analysis}

\begin{figure*}[htbp]
  \centering
  \makebox[\textwidth][c]{\includegraphics[width=1.0\textwidth]{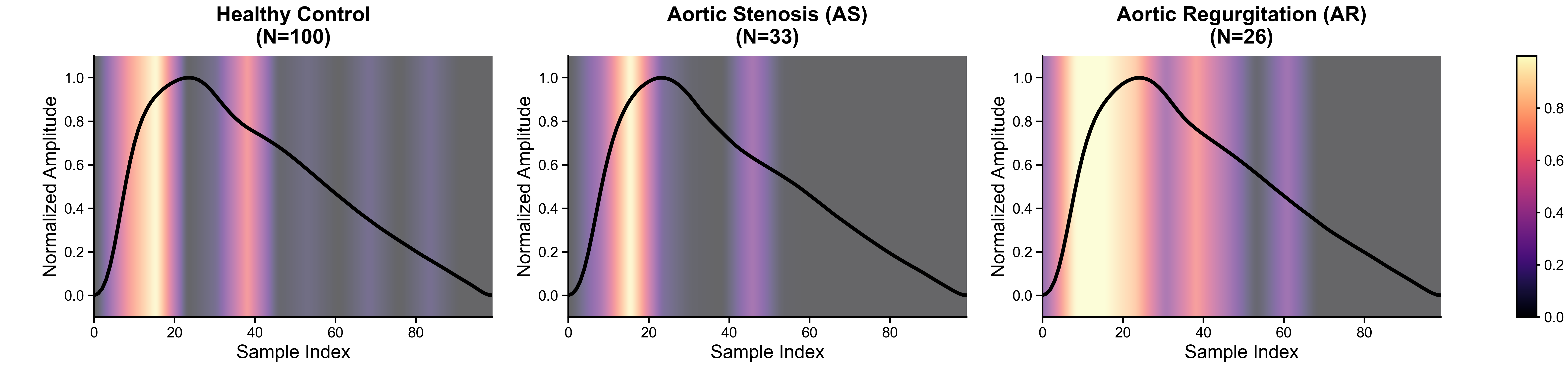}}
  \caption{Grad-CAM visualization of model attention across representative PPG waveforms from healthy controls, AS, and AR groups. Lighter colors indicate higher Grad-CAM activation and greater contribution to the model prediction. Across all groups, attention was concentrated within physiologically meaningful regions of the pulse contour.}
  \label{fig:grad_cam}
\end{figure*}

To qualitatively examine the interpretability of the learned representations, we visualized Grad-CAM activation patterns across representative waveforms from healthy controls, AS, and AR groups. As shown in Figure~\ref{fig:grad_cam}, attention across all waveform categories was concentrated within physiologically meaningful regions of the pulse contour rather than being diffusely distributed across the entire waveform, suggesting that the model primarily relies on informative hemodynamic components of the PPG signal when generating predictions~\cite{elgendi2012analysis}. Distinct attention distributions were nevertheless observed among the three waveform categories.

Within this shared attention framework, the AS group exhibited a more localized attention pattern than that observed in healthy controls, with attention concentrated within two relatively confined waveform regions. These regions span phases associated with systolic ejection and the subsequent propagation of the pulse wave through the arterial system, suggesting relatively greater attention to waveform components linked to ventricular output and cardiovascular flow dynamics. Alterations in these physiological processes are well-recognized features of AS-associated hemodynamics~\cite{abbas2019hemodynamic}. Although vascular stiffness and age-related arterial changes also influence pulse-wave morphology~\cite{hungerford2023interpretation}, the observed concentration of attention suggests that the model relies on waveform regions carrying information relevant to the hemodynamic consequences of AS.

The AR group exhibited a broader attention distribution than that observed in healthy controls, with attention extending across a larger portion of the waveform from the systolic phase into the descending phase. Because PPG reflects pulsatile peripheral blood volume dynamics, chronic regurgitant hemodynamics and widened pulse pressure can influence pulse morphology throughout multiple phases of the cardiac cycle~\cite{piccioli2023cardiac}. Sustained attention within the descending portion of the waveform further suggests that the model relies on information distributed across multiple physiologically relevant phases of the pulse contour rather than on a narrowly localized waveform region. Together, these observations indicate that AR-related predictions are informed by waveform characteristics spanning a broad range of pulse-wave dynamics.

The Healthy Control group exhibited attention concentrated within several major waveform regions while maintaining lower-level attention across a broader portion of the pulse contour. Compared with the more localized attention pattern observed in the AS group, this distribution suggests that the model incorporates information from multiple waveform segments when identifying non-pathological waveforms, rather than relying predominantly on a small number of highly focused regions. Overall, these visualization results suggest that the representations learned by PiLA are aligned with physiologically meaningful waveform components while exhibiting different attention distributions across waveform categories.

\section*{DISCUSSION}
The management of cardiovascular diseases is increasingly shifting from traditional symptom-driven models toward more proactive screening strategies enabled by wearable technologies~\cite{pedroso2025leveraging, wong2024global}. Such opportunistic screening approaches have demonstrated substantial potential for improving the identification of under-recognized cardiovascular conditions while offering favorable cost-effectiveness~\cite{chen2022cost}. In this context, our study demonstrates the feasibility of using large-scale PPG data to develop a risk-enrichment framework for AVD screening using the proposed PiLA framework. Constrained by the scarcity of clinically labeled data, we adopted a physiology-guided self-supervised learning strategy to leverage large-scale unlabeled PPG signals for representation learning. Our results show that PiLA improves screening performance for both AS and AR using only limited labeled samples. These findings support the potential utility of physiology-guided representation learning for low-cost cardiovascular screening and risk-enrichment applications in data-limited settings.

Our comparative experiments further highlight the importance of physiologically meaningful inductive bias in medical AI. While physiological feature clustering improved performance relative to purely supervised learning, its gains remained substantially smaller than those achieved by PiLA. This observation suggests that not all forms of physiological knowledge contribute equally to representation learning. In contrast, the effectiveness of PiLA may stem from incorporating physiological priors that are more closely aligned with clinically recognized hemodynamic characteristics of AVD, such as pulsus tardus. Such targeted physiological guidance may help the model focus on disease-associated waveform patterns despite substantial inter-individual physiological variation. These findings suggest that, in label-limited settings, appropriately designed domain-specific priors may provide a useful strategy for improving representation learning from physiological signals.

From a physiological perspective, our findings further support the relevance of PPG as a non-invasive hemodynamic signal for cardiovascular screening. By reflecting peripheral blood volume dynamics, PPG may capture waveform alterations associated with cardiovascular hemodynamic changes related to valvular dysfunction. Longitudinal analyses further demonstrated that individuals identified as high-risk by PiLA were associated with a higher incidence of subsequently recorded AVD diagnoses, with risk stratification remaining significant after adjustment for traditional cardiovascular risk factors such as age, hypertension, and BMI through multivariable Cox regression and Propensity Score Matching analyses. These findings suggest that the learned representations capture physiological waveform patterns associated with clinically relevant hemodynamic variation beyond routine demographic and cardiovascular risk factors. Collectively, these results support the potential utility of PPG-based screening and risk-enrichment strategies for identifying individuals who may benefit from further cardiovascular evaluation prior to future clinically recognized AVD events.

Despite the rapidly growing availability of wearable physiological signals, obtaining high-quality clinical labels remains a major bottleneck in medical AI~\cite{azizi2023robust}. This challenge is particularly pronounced in PPG analysis, where substantial inter-subject physiological variability limits the effectiveness of purely supervised learning under label-scarce conditions~\cite{ghorbani2023self}. Against this backdrop, increasing attention has been directed toward data-centric AI strategies that aim to better leverage the latent information embedded within unlabeled data~\cite{majeed2024data, sun2026review}. Recent advances in large-scale physiological representation learning have further demonstrated the potential of self-supervised pretraining on wearable and biosignal data for applications such as activity recognition, sleep analysis, and disease prediction~\cite{yuan2024self, narayanswamy2025scaling, thapa2026sleepfm}. However, most existing approaches rely primarily on generic representation objectives designed to capture broad temporal or behavioral structure. While highly effective for scalable representation learning, such objectives may not always be optimally aligned with the subtle hemodynamic waveform variations relevant to valvular heart disease. In this context, our findings suggest that the effectiveness of physiology-guided self-supervised learning may arise not only from large-scale pretraining itself, but also from the physiological alignment between the pretraining objective and downstream hemodynamic variation. Rather than relying solely on generic signal-level invariances, PiLA incorporates clinically motivated physiological priors to guide representation learning toward disease-relevant waveform characteristics associated with AVD. Experimental results further suggest that large-scale pretraining may help improve robustness to imperfect proxy supervision and inter-individual physiological variability.

Nevertheless, several limitations should be acknowledged. First, the rule-based generation of proxy supervision remains relatively heuristic and may introduce label noise. Second, although individuals diagnosed within one year after PPG acquisition were included to reduce under-ascertainment caused by delayed clinical recognition of AVD, the absence of contemporaneous echocardiographic confirmation limits definitive interpretation of disease status at the time of signal acquisition. While the diagnosis-defined labels were derived from clinically recorded diagnoses and self-reported medical history within the routine healthcare system, they should not be considered equivalent to contemporaneous echocardiography-confirmed disease status. Third, the current framework was developed using short-duration averaged PPG waveforms acquired under controlled conditions and therefore cannot be directly generalized to continuous real-world wearable monitoring scenarios. Fourth, the study was conducted within the UK Biobank, whose participants are predominantly of European ancestry and consist largely of middle-aged and older adults. Consequently, the cohort may not fully represent the demographic, ethnic, and cardiovascular diversity of the general population. External validation in more diverse populations, age groups, and acquisition settings will therefore be necessary to assess generalizability.

Taken together, our findings support the feasibility of physiology-guided self-supervised learning for leveraging unlabeled physiological signals in data-limited cardiovascular screening settings. Beyond AVD screening itself, this framework may provide a useful direction for integrating clinically motivated physiological priors into representation learning for wearable cardiovascular AI.

\section*{RESOURCE AVAILABILITY}

\subsection*{Lead Contact}
Further information and requests for resources should be directed to and will be fulfilled by the lead contact, Shenda Hong (hongshenda@pku.edu.cn).

\subsection*{Materials Availability}
This study did not generate new unique reagents.

\subsection*{Data and Code Availability}
\begin{itemize}
    \item \textbf{Data:} This study used data from the UK Biobank. Access to UK Biobank data is not publicly available and requires an approved application through the UK Biobank (\url{https://www.ukbiobank.ac.uk/}).
    \item \textbf{Code:} All original code for the PiLA framework has been deposited at \url{https://github.com/EZAIJ/PiLA-PPG} and \url{https://github.com/PKUDigitalHealth/PiLA-PPG} are publicly available as of the date of publication.
    \item \textbf{Additional Info:} Any further inquiries regarding the \textit{implementation details or model architecture} are available from the lead contact upon request.
\end{itemize}

\section*{METHODS}

\subsection{Cohort Construction and Endpoint Definition}

The data used in this study were derived from the UK Biobank. PPG signals were extracted from Field 4205, acquired using the PulseTrace PCA2 device. The UK Biobank provides each recording as a standardized 100-point averaged pulse waveform derived from a short acquisition period. For participants with multiple records, only the baseline acquisition was retained, such that each participant contributed a single baseline waveform.

Positive cases were identified using ICD-10 diagnostic codes (AS: I35.0/I06.0; AR: I35.1/I06.1) and self-reported medical history (UK Biobank Field 20002). Because AVD is a chronic and often under-recognized condition, the date of clinical diagnosis may lag behind the actual presence of disease. Therefore, diagnoses recorded within one year after PPG acquisition were also included in the positive case definition. This short post-acquisition window was intended to capture clinically unrecognized but likely prevalent disease at the time of PPG measurement, rather than to define a future prediction endpoint. Participants with a history of aortic valve intervention were excluded based on OPCS-4 procedural codes (K26, K30, K35.2). Controls were sampled from participants without recorded AS or AR within the baseline ascertainment window.

A downstream labeled cohort was then constructed by combining all identified AS- and AR-positive participants with sampled controls. This downstream cohort was partitioned at the subject level into training, validation, and independent test subsets. To maintain strict participant-level separation for unbiased final evaluation, all participants assigned to the independent test set were excluded from the pretraining cohort. The remaining participants were eligible for physiology-guided pretraining, during which no AS/AR diagnostic labels were used. Thus, overlap between pretraining and downstream fine-tuning was permitted only for the training and validation subsets and only in an unlabeled/proxy-supervised manner, while the independent test set remained completely unseen during all training stages.

Unless otherwise stated, analyses requiring alternative evaluation cohorts were performed by repeating the complete model development pipeline. Participants assigned to each evaluation cohort were excluded from physiology-guided pretraining prior to model development to maintain strict separation between model development and evaluation.

\subsection{Formal Problem Definition}

\label{sec:problem_definition}

The primary challenge addressed in this research is the robust screening of AVD from Photoplethysmography signals under the realistic constraint of limited availability of high-quality clinical labels.

To tackle this challenge, we formulate the problem within the Physiology-Guided Self-Supervised Learning paradigm, in which physiology-derived proxy supervision is used to facilitate large-scale representation learning prior to downstream transfer learning.

Specifically, we consider two distinct types of datasets:

\begin{itemize}

    \item \textit{A large-scale dataset without AVD-specific diagnostic labels}, denoted as $\mathcal{D}_U = \{ \mathbf{x}_i \}_{i=1}^{N_U}$. Each sample $\mathbf{x}_i \in \mathbb{R}^{L}$ is a pre-processed PPG time-series segment of length $L$. Crucially, this dataset is augmented with physiological metadata derived from pulse wave analysis, which enables the generation of proxy labels for our self-supervised pretext task.

    \item \textit{A small-scale target dataset with diagnosis-defined AVD labels}, denoted as $\mathcal{D}_L = \{ (\mathbf{x}_j, \mathbf{y}_j) \}_{j=1}^{N_L}$, where the number of samples is significantly smaller than in the unlabeled set ($N_L \ll N_U$). The label for each sample is a binary vector $\mathbf{y}_j = [y_j^{(\text{AS})}, y_j^{(\text{AR})}] \in \{0, 1\}^2$, indicating the presence or absence of AS and AR.

\end{itemize}

Our central objective is to learn a powerful deep learning model $F: \mathbb{R}^{L} \to [0, 1]^2$, which is composed of a feature encoder $f_\theta(\cdot)$ and a classifier $g_\phi(\cdot)$, such that $F(\cdot) = g_\phi(f_\theta(\cdot))$. 

The encoder $f_\theta$, parameterized by $\theta$, is primarily pre-trained on the large-scale dataset $\mathcal{D}_U$ through a bespoke physiology-guided self-supervised proxy task. This pre-training stage aims to learn physiologically relevant waveform representations from large-scale unlabeled PPG data before downstream adaptation to AVD screening tasks.

Subsequently, the entire model is fine-tuned on the small labeled dataset $\mathcal{D}_L$ to achieve accurate and reliable screening for both AS and AR.

\subsection{Overall Architecture}

\label{sec:overall_architecture}

To address the challenge of AVD screening in a label-scarce environment, we developed the PiLA framework as the implementation of the proposed PG-SSL strategy. Serving as the concrete implementation of the PG-SSL paradigm, the core of this framework lies in the integration of two core modules:

\begin{enumerate}

    \item \textbf{PPG Morphological Phenotyping for AVD:} An analysis module responsible for systematically identifying and quantifying abnormal PPG patterns associated with the pathological states of AS and AR from a physiological perspective.

    \item \textbf{Physiology-Guided Self-Supervised Learning:} An advanced deep learning framework that leverages the physiological patterns defined by the first module as prior knowledge. It addresses the model learning problem through a pipeline of self-supervised pre-training and subsequent gated knowledge-fusion fine-tuning.

\end{enumerate}

\begin{figure*}[htbp]
    \centering
    \includegraphics[width=0.85\textwidth]{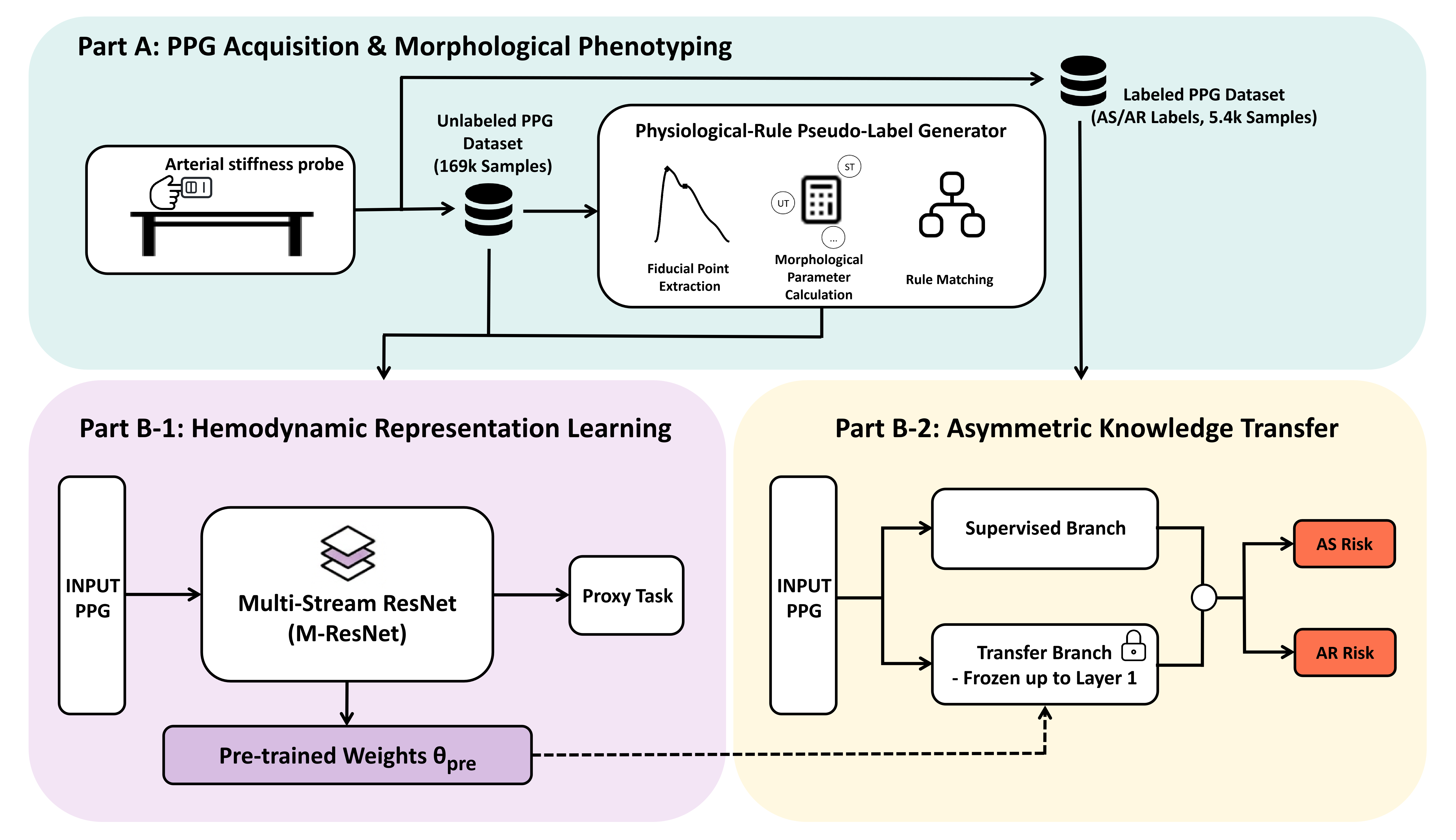} 
    \caption{Overview of the PiLA framework. Part A illustrates PPG acquisition from the UK Biobank and construction of physiology-derived pseudo-labels through morphological phenotyping. Part B-1 presents physiology-guided self-supervised pretraining for hemodynamic representation learning using the proposed M-ResNet encoder. Part B-2 shows the downstream fine-tuning stage, in which pretrained representations are transferred through an asymmetric knowledge transfer framework for AS and AR risk prediction.
}
    \label{fig:overall_framework}
\end{figure*}

Our overall workflow, as illustrated in Figure~\ref{fig:overall_framework}, clearly demonstrates how these two modules collaborate within a three-stage process encompassing data collection, pre-training, and fine-tuning. This integrated process enables end-to-end prediction of AVD from raw PPG signals.

\FloatBarrier

\subsection{PPG Morphological Phenotyping for AVD}

We first established PPG morphological phenotypes associated with the distinct hemodynamic alterations of AS and AR. AS and AR leave characteristic imprints on the arterial pulse wave through distinct hemodynamic mechanisms. These morphological changes, originating from the central arterial pressure wave, can be non-invasively captured by peripheral PPG signals~\cite{millasseau2006contour}.

In AS, the calcified aortic valve restricts outflow, leading to a significant increase in left ventricular ejection impedance. The blood flow is forced through a narrowed orifice, which not only prolongs the ejection time but also, due to the sustained pressure gradient, creates a characteristic pulse with a slow-rising upstroke and a delayed peak, known as \textit{pulsus tardus}~\cite{moxham2003understanding}. In severe AS, a secondary shock wave at the stenotic valve can also cause a distinct notch on the upstroke, forming an \textit{anacrotic pulse}~\cite{fleming1957mechanism}.

In AR, the incomplete closure of the aortic valve allows blood to regurgitate back into the left ventricle during diastole. This compels the heart to eject an augmented stroke volume under a hyperdynamic state during systole. The rapid ejection of a large volume of blood against a low-resistance periphery creates a steep upstroke, while the diastolic pressure plummets due to the backflow. Together, these effects result in a widened pulse pressure and a rapidly rising and falling pulse known as the WHP~\cite{suvarna2008watson, boiteau1964upstroke}. From a hemodynamic perspective, these abnormal patterns precisely capture the core pathophysiological features of AS and AR across the dimensions of ejection timing, velocity, and flow impact morphology.

Based on these hemodynamic characteristics, we next developed a framework to quantify AVD-related PPG morphological phenotypes for large-scale proxy supervision. Our framework is inspired by the methodology of Meghraoui and Yoshioka et al.~\cite{meghraoui2024classifying,yoshioka2010do}. 

We used UK Biobank fields 4194-0.0 (heart rate), 4198-0.0 (peak position), 4199-0.0 (notch position), and 4200-0.0 (shoulder position) to quantify pulse wave morphology. From these fields, we computed the upstroke time (UT) and systolic time (ST) by multiplying the respective peak and notch positions by $\Delta t$, with $\Delta t$ calculated as 60 divided by the heart rate and by 100 to represent the duration of a single sampling interval. The pulse peak-to-shoulder interval (PPT) was defined as the difference between peak and shoulder positions when a shoulder was detected, and the shoulder position itself (OPT) was taken as the corresponding time of the shoulder.

Anacrotic pulses were defined by the presence of a shoulder with PPT exceeding OPT. Pulsus Tardus was defined for waveforms with UT exceeding 0.156 seconds and a UT/ST ratio exceeding 0.5. Water-Hammer Pulses were defined for waveforms with UT below 0.11 seconds and a UT/ST ratio below 0.34. These rule-derived phenotypes were generated from UK Biobank pulse wave analysis parameters rather than directly provided diagnostic labels.

The proposed Arterial Pulse Pattern Recognition (APPR) task was formulated as a multi-task binary proxy supervision problem rather than a mutually exclusive multiclass classification task. Each waveform was independently assigned three binary pseudo-labels corresponding to Anacrotic, Tardus, and WHP morphological patterns. Collectively, these binary proxy labels served as physiology-guided supervision for large-scale pretraining.

Although the aforementioned pathological patterns are theoretically distinct, their reliable identification in real-world peripheral PPG signals is complicated by a multitude of confounding physiological factors. A key confounder is arterial stiffness, which increases with age. Increased stiffness accelerates the return of the reflected pulse wave, causing it to superimpose on the systolic phase, which can lead to a forward shift of the peak and alterations in the dicrotic notch morphology~\cite{karimpour2023photoplethysmography}. Some of these age-related physiological changes can mimic the features of AR, leading to significant confusion. Furthermore, the PPG waveform is dynamically modulated by other physiological states, such as heart rate and blood pressure, meaning that any heuristic rules based on fixed thresholds will inevitably be noisy and uncertain. Rather than treating these rule-derived phenotypes as deterministic classifiers, we leveraged them to construct a structured pretext task for self-supervised learning. On such a massive dataset, a deep neural network, in its effort to minimize the global loss, must learn beyond simply fitting ambiguous boundary cases. Instead, it is compelled to learn the underlying morphological features associated with the vast majority of typical physiological waveform patterns rather than overfitting to ambiguous boundary cases. The scale of the dataset provides strong statistical robustness to the proxy task, encouraging the model to learn stable waveform representations that generalize beyond ambiguous individual cases. Consequently, these morphology-derived phenotypes served as physiology-guided proxy supervision for large-scale pretraining.

\subsection{Neural Architecture of PiLA}

\label{sec:pg-ssl_framework}

\begin{figure*}[htbp]
    \centering
    \includegraphics[width=0.85\textwidth]{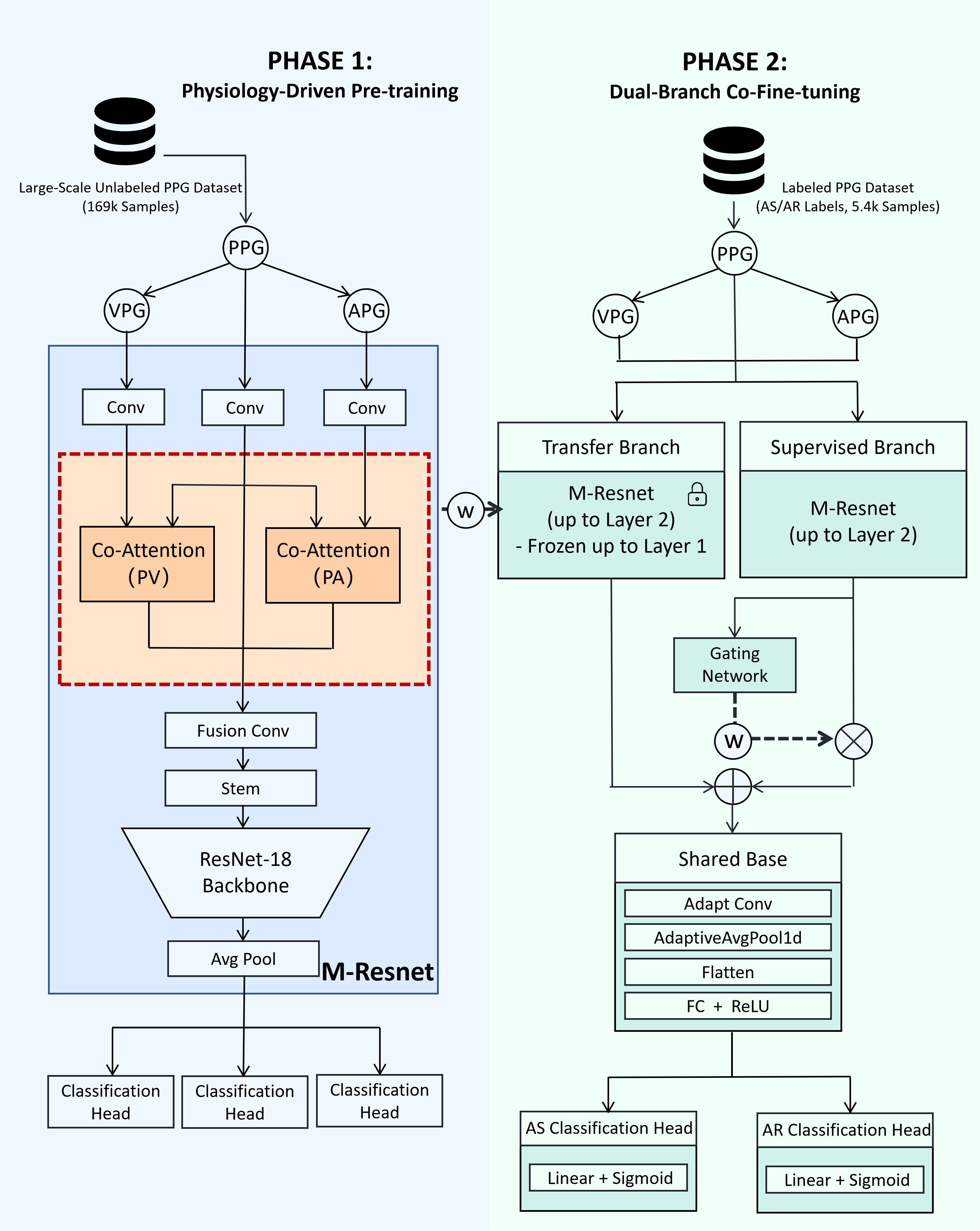} 
    \caption{Neural architecture of the PiLA framework. Phase 1 (left) depicts the physiology-guided pretraining architecture, where the proposed M-ResNet processes multi-stream PPG, VPG, and APG inputs with a co-attention mechanism. Phase 2 (right) illustrates the downstream dual-branch fine-tuning architecture. The Transfer Branch loads partially frozen pretrained weights, whereas the Supervised Branch is trained from scratch. A gating network integrates features from both branches before the shared classification backbone for AS and AR screening.}
    \label{fig:model_architecture}
\end{figure*}

To leverage large-scale PPG data without AVD-specific diagnostic labels for physiology-guided representation learning under the constraint of limited clinical labels, and to generalize these features to downstream classification tasks, we developed the neural architecture of PiLA. As illustrated in Figure~\ref{fig:model_architecture}, this framework comprises two core stages: self-supervised pre-training based on multi-stream enhancement, and fine-tuning based on asymmetric feature complementation.

The first stage aims to learn physiologically relevant waveform representations from the large-scale dataset $\mathcal{D}_U$ through the APPR proxy task. We utilize the PPG morphological phenotypes defined in the previous section as pseudo-labels to construct an \textit{Arterial Pulse Pattern Recognition} proxy task, encouraging the model to learn physiologically relevant waveform representations that generalize across heterogeneous physiological backgrounds.

In terms of architecture design, we construct a feature encoder named Multi-Stream ResNet (M-ResNet). Although the ResNet1D-18 has been widely validated, standard convolutional networks tend to smooth out high-frequency details during downsampling, whereas the core pathological features of AS and AR often manifest as subtle deformations occurring in extremely short instants. To overcome this limitation, we modified the front-end input stem of the ResNet by designing a \textit{multi-modal co-input module}. Specifically, in addition to the raw PPG signal, we explicitly introduce its first derivative (VPG) and second derivative (APG) as parallel input channels. This design leverages the high sensitivity of derivative signals to waveform inflection points, encouraging the encoder to jointly capture waveform morphology and derivative-based temporal dynamics between the raw waveform and instantaneous velocity and acceleration at the initial stage of feature extraction. Upon completion of training, we save the optimal weights $\theta_{pre}$ as the initialization for subsequent tasks.

The second stage transfers the pretrained encoder to downstream AS and AR screening. Given the scarcity of clinical labels in $\mathcal{D}_L$, we design a \textit{dual-branch asymmetric fusion} architecture. This architecture contains two parallel branches, both utilizing the M-ResNet structure described above. To preserve and integrate mid-level morphological features potentially relevant to AVD screening, we choose to perform feature integration at the intermediate layer (specifically after Layer 2) of the network rather than at the end.

To coordinate generality with specificity, we propose an asymmetric feature complementation strategy. Specifically, we freeze the front-end multi-modal module of the Transfer Branch to preserve low-level waveform representations, while transmitting its output feature $h_{ssl}$ directly as a foundational representation. In contrast, the Supervised Branch is initialized from scratch to capture the distribution characteristics of the specific dataset, denoted as $h_{task}$. To prevent this branch from introducing random noise due to the small data volume, we introduce a lightweight gating network as a correlation filter to dynamically suppress low-confidence features. Finally, the fused feature is obtained via the formula $h_{fused} = h_{ssl} + (h_{task} \odot w_{gate})$. This mechanism ensures that the model retains the complete general representation while only adding filtered, task-specific waveform features as an increment.

Finally, the fused features are sent into a shared backend network for deep integration and are used to predict the probabilities of AS and AR respectively through two independent linear classification heads. The trainable components of the model were optimized jointly during the fine-tuning phase using a Weighted Binary Cross-Entropy Loss to mitigate the impact of positive-negative sample imbalance.

\FloatBarrier

\subsection{Clinical Confounding and Error Characterization}

To evaluate the extent to which model performance could be explained by measurable demographic and clinical factors, we conducted a clinical-feature reference experiment, a covariate-balancing analysis, and an error characterization analysis.

For the clinical-feature reference experiment, a LightGBM classifier was trained using routinely available demographic characteristics, vital signs, and comorbidity history, including age, sex, body mass index (BMI), heart rate, systolic blood pressure, diastolic blood pressure, smoking status, diabetes mellitus, chronic kidney disease, and cardiovascular disease history. Gradient boosting was selected because it can flexibly model nonlinear relationships among conventional clinical risk factors and has been widely used as a strong tabular-data baseline.

For the propensity-score balancing analysis, a covariate-balanced downstream cohort was constructed prior to model fine-tuning. Propensity scores were estimated using logistic regression based on age, sex, BMI, arterial stiffness index (ASI), hypertension, diabetes mellitus, hyperlipidemia, chronic kidney disease, coronary artery disease/myocardial infarction, atrial fibrillation, and stroke history. Missing values were imputed using median imputation and variables were standardized before propensity-score estimation. Control participants were selected through propensity-score stratified sampling to construct a covariate-balanced cohort while preserving the overall propensity-score distribution of the case group. Covariate balance was evaluated using standardized mean differences, with an absolute SMD below 0.1 considered acceptable balance. The resulting covariate-balanced cohort was subsequently used for downstream model fine-tuning and evaluation.

For error characterization, participants in the independent test cohort were ranked according to model-predicted risk scores separately for AS and AR. Individuals within the top 10\% of model-derived risk scores were designated as the high-risk group. Based on disease labels and risk-group assignment, participants were categorized into TP, FN, FP, and TN groups, and major demographic and clinical characteristics were compared between groups. Continuous variables were compared using Welch’s t-test, whereas categorical variables were compared using Fisher’s exact test.

\subsection{AS--AR Subtype Discrimination Analysis}

To evaluate whether PiLA captured subtype-specific information beyond general AVD-related signals, we performed an AS--AR subtype discrimination analysis restricted to participants with isolated AS or isolated AR, excluding mixed AVD cases and non-AVD controls.

Two complementary analyses were performed. First, the difference between the AS and AR prediction probabilities generated by PiLA was used as a subtype discrimination score. Second, embeddings extracted immediately before the task-specific output heads were used to train an L2-regularized logistic regression linear probe to distinguish isolated AS from isolated AR. The linear probe was trained on the training set and evaluated on the held-out test set.

\subsection{Temporal Sensitivity Analysis}

To evaluate the temporal sensitivity of the proposed framework across different diagnostic horizons, we performed a diagnostic-horizon analysis based on the interval between PPG acquisition and the first recorded clinical diagnosis of AS or AR. For each disease, participants were stratified into four independent evaluation cohorts according to the time-to-diagnosis interval: 1--3 years, 3--5 years, 5--10 years, and greater than 10 years after PPG acquisition.

The trained model was independently evaluated within each diagnostic-horizon cohort. Receiver operating characteristic curves, area under the ROC curve, and sensitivity at a fixed specificity of 70\% were calculated separately for each interval subgroup. To estimate statistical uncertainty, 95\% confidence intervals for AUROC values were calculated using bootstrap resampling with 1000 iterations, and temporal performance trends across diagnostic horizons were subsequently visualized.

\subsection{Long-term Risk Stratification and Independent Prognostic Value}

To evaluate whether the learned PPG representations retained longitudinal risk-associated information beyond cross-sectional disease identification, we constructed separate incident follow-up cohorts for AS and AR from the baseline PPG population. Consistent with the downstream screening task, participants experiencing an AS- or AR-related endpoint event before or within one year after PPG acquisition were considered prevalent cases and excluded from the corresponding incident analyses. Participants with a history of aortic valve surgery or intervention before baseline were also excluded. Aortic valve surgery or intervention was defined as including valve repair, replacement, and valvotomy procedures identified from linked hospital procedure records. Follow-up started at the baseline PPG assessment date and continued until the first incident AS or AR diagnosis, AS- or AR-related death, or aortic valve surgery or intervention. Participants without an outcome event were censored at non-related death, loss to follow-up, or the administrative end of linked healthcare follow-up (31 May 2022), whichever occurred first. Outcome ascertainment was based on linked healthcare records and death registry data rather than subsequent UK Biobank assessment visits. The resulting incident follow-up cohorts comprised 166,259 participants with 1,563 incident AS events and 166,260 participants with 1,143 incident AR events during follow-up, respectively.

To evaluate risk stratification under a more balanced clinical setting, separate propensity-score matched (PSM) cohorts were constructed for AS and AR within the incident follow-up population. Propensity scores were estimated using age, sex, BMI, arterial stiffness index, hypertension, diabetes mellitus, hyperlipidemia, chronic kidney disease, coronary artery disease/myocardial infarction, atrial fibrillation, and stroke history. Missing values for continuous variables were imputed using median imputation, and corresponding missing-indicator variables were included during propensity-score estimation. Incident AS and AR cases were matched to controls using 1:4 nearest-neighbor propensity-score matching without replacement and a caliper width of 0.2 standard deviations of the logit propensity score.

Covariate balance was assessed using standardized mean differences, with an absolute SMD below 0.1 considered acceptable balance. The final matched cohorts comprised 7,815 participants for AS (1,563 cases and 6,252 controls) and 5,715 participants for AR (1,143 cases and 4,572 controls).

Kaplan--Meier survival analyses were performed in the matched cohorts using both quartile-based risk stratification and a binary high-risk screening strategy defined by the top 25\% of predicted risk scores. Survival differences between groups were assessed using log-rank tests, and hazard ratios were estimated using Cox proportional hazards regression. For multiple survival comparisons, $P$ values were adjusted using the Benjamini--Hochberg false discovery rate procedure.

To further evaluate the independent prognostic value of the AI-derived waveform representations, multivariable Cox proportional hazards regression analyses were performed in the full incident follow-up cohort. Model-predicted probabilities were standardized and entered as continuous predictors; hazard ratios therefore represent the relative hazard per one-standard-deviation increase in the PiLA score. Three progressively adjusted models were constructed: Model 1 adjusted for age, sex, and BMI; Model 2 additionally adjusted for systolic blood pressure, diastolic blood pressure, smoking status, and arterial stiffness index; and Model 3 further adjusted for hypertension, diabetes, hyperlipidemia, chronic kidney disease, coronary artery disease/myocardial infarction, atrial fibrillation, and stroke.

\subsection{Subgroup Analysis and Physiological Heterogeneity}

To evaluate the robustness of the proposed framework across heterogeneous physiological backgrounds, subgroup analyses were performed across predefined demographic and physiological strata, including sex, age, BMI, blood pressure status, arterial stiffness, diabetes history, and hypertension history. Subgroup definitions were derived from corresponding UK Biobank clinical variables using clinically motivated threshold-based categorization.

Because the number of AS and AR cases in the independent test set was limited, subgroup analyses were conducted using out-of-fold predictions generated from five-fold cross-validation on the downstream cohort. Receiver operating characteristic analysis was performed independently for each subgroup, and subgroup-specific area under the ROC curve values with 95\% confidence intervals were estimated using bootstrap resampling with 1000 iterations.

To further assess subgroup-level performance consistency, statistical differences between paired subgroup AUROC distributions were evaluated using Welch’s two-sample t-test applied to the bootstrap AUROC distributions. Forest plots were subsequently generated to visualize subgroup-specific performance estimates and confidence intervals relative to the overall cohort performance.

\subsection{Model Interpretation Analysis}

To evaluate the physiological relevance and interpretability of the learned representations, we applied one-dimensional Gradient-weighted Class Activation Mapping (1D Grad-CAM) to visualize waveform regions contributing to model predictions. Grad-CAM maps were generated separately for the AS and AR prediction tasks using samples from the independent test cohort. Representative activation maps were computed for AS, AR, and healthy control groups and overlaid onto the corresponding averaged PPG waveforms for visualization. To improve visualization stability and reduce individual variability, group-level Grad-CAM maps were generated by averaging activation patterns across multiple subjects within each category. These visualizations were used to qualitatively assess whether the model attended to physiologically plausible waveform regions associated with different valvular hemodynamic patterns.

\subsection{Ethics statement}

The study was conducted in accordance with the Declaration of Helsinki. Ethical approval for the UK Biobank study was obtained from the North West-Haydock Research Ethics Committee (REC reference: 21/NW/0157). All participants provided written informed consent at recruitment for their data to be used in health-related research. This study was conducted under UK Biobank application number 90018.

\section*{ACKNOWLEDGMENTS}
This work was supported in part by the National Natural Science Foundation of China under Grant 62102008, Grant 62202332, Grant 62376197, Grant 62020106004 and Grant 92048301; in part by the CCF-Zhipu Large Model Innovation Fund (CCF-Zhipu202414); in part by the CCF-Tencent Rhino-Bird Open Research Fund (CCF-Tencent RAGR20250108); in part by the Tianjin Science and Technology Program under Grant 23JCYBJC00360; in part by the Research Project of Peking University in the State Key Laboratory of Vascular Homeostasis and Remodeling (2025-SKLVHR-YCTS-02); in part by the PKU-OPPO Fund (BO202301, BO202503). 

\section*{DECLARATION OF INTERESTS}
The authors declare no competing interests.

\bibliography{reference} 

\end{document}